\documentclass{emulateapj}

\usepackage{graphicx}
\usepackage{times}
\usepackage{subfigure}
\usepackage{color}
 \usepackage{hyperref}	
 
\renewcommand\appendix{\par
  \setcounter{section}{0}
  \setcounter{subsection}{0}
  \setcounter{figure}{0}
  \setcounter{table}{0}
  \renewcommand\thesection{Appendix \Alph{section}}
  \renewcommand\thefigure{\Alph{section}\arabic{figure}}
  \renewcommand\thetable{\Alph{section}\arabic{table}}
}

\newcommand{\CII}{\hbox{{\rm C}\kern 0.1em{\sc ii}}}
\newcommand{\CIV}{\hbox{{\rm C}\kern 0.1em{\sc iv}}}
\newcommand{\Fe}{\hbox{{\rm Fe}}}
\newcommand{\FeI}{\hbox{{\rm Fe}\kern 0.1em{\sc i}}}
\newcommand{\FeII}{\hbox{{\rm Fe}\kern 0.1em{\sc ii}}}
\newcommand{\Ti}{\hbox{{\rm Ti}}}
\newcommand{\Si}{\hbox{{\rm Si}}}
\newcommand{\SiII}{\hbox{{\rm Si}\kern 0.1em{\sc ii}}}

\newcommand{\AlII}{\hbox{{\rm Al}\kern 0.1em{\sc ii}}}
\newcommand{\AlIII}{\hbox{{\rm Al}\kern 0.1em{\sc iii}}}
\newcommand{\NiII}{\hbox{{\rm Ni}\kern 0.1em{\sc ii}}}

\newcommand{\Cr}{\hbox{{\rm Cr}}}
\newcommand{\CrII}{\hbox{{\rm Cr}\kern 0.1em{\sc ii}}}

\newcommand{\Mn}{\textsc{{\rm Mn}}}
\newcommand{\MnII}{\textsc{{\rm Mn}\kern 0.1em{\sc ii}}}
\newcommand{\TiII}{\textsc{{\rm Ti}\kern 0.1em{\sc ii}}}
\newcommand{\Zn}{\textsc{{\rm Zn}}}
\newcommand{\ZnII}{\textsc{{\rm Zn}\kern 0.1em{\sc ii}}}

\newcommand{\NeV}{\textsc{[{\rm Ne}\kern 0.1em{\sc v}}]}
\newcommand{\NII}{\textsc{[{\rm N}\kern 0.1em{\sc ii}}]}
\newcommand{\OIII}{\textsc{[{\rm O}\kern 0.1em{\sc iii}}]}
\newcommand{\OII}{\textsc{[{\rm O}\kern 0.1em{\sc ii}}]}
\newcommand{\OI}{\textsc{{\rm O}\kern 0.1em{\sc i}}}
\newcommand{\MgI}{\textsc{{\rm Mg}\kern 0.1em{\sc i}}}
\newcommand{\MgII}{\textsc{{\rm Mg}\kern 0.1em{\sc ii}}}
\newcommand{\HH}{\textsc{{\rm H}}}
\newcommand{\HI}{\textsc{{\rm H}\kern 0.1em{\sc i}}}
\newcommand{\HII}{\textsc{{\rm H}\kern 0.1em{\sc ii}}}
\newcommand{\lya}{\textsc{{\rm Ly}\kern 0.1em$\alpha$}}
\newcommand{\Ly}{\textsc{{\rm Ly}\kern 0.1em$\alpha$}}
\newcommand{\Ha}{\textsc{{\rm H}\kern 0.1em$\alpha$}}
\newcommand{\Hb}{\textsc{{\rm H}\kern 0.1em$\beta$}}
\newcommand{\Hg}{\textsc{{\rm H}\kern 0.1em$\gamma$}}
\newcommand{\SII}{\hbox{[{\rm S}\kern 0.1em{\sc ii}}]}

\newcommand{\fdlam}{erg~s$^{-1}$~cm$^{-2}$~\AA$^{-1}$}

\newcommand{\flux}{erg~s$^{-1}$~cm$^{-2}$}

\newcommand{\mpy}{\hbox{M$_{\odot}$~yr$^{-1}$}}

\newcommand{\msun}{\hbox{M$_{\odot}$}}
\newcommand{\cmsq}{\hbox{cm$^{-2}$}}
\newcommand{\NHI}{\ensuremath{N_\textsc{h\scriptsize{\,i}}}}
\newcommand{\NH}{\hbox{$N_{\rm H}$}}

\newcommand{\kpc}{\hbox{$h^{-1}$~kpc}}

\newcommand{\kms}{km~s$^{-1}$}
\newcommand{\nn}{\nonumber}
\newcommand{\EW}{\hbox{$W_{\rm r}^{\lambda2796}$}}

\newcommand{\redshift}{\hbox{0.9096 $\pm$ 0.0001}}

\newcommand{\logZ}{\log Z/Z_{\odot}}

\newcommand{\sfr}{$4.7\pm2.0$~\mpy}

\newcommand{\galpak}{{GalPaK$^{\rm 3D}$}}

\slugcomment{Received; Accepted; Draft version \today}

\shorttitle{Accretion in star-forming galaxies}
\shortauthors{Bouch\'e N. et al.}

\begin{document}
\title{Possible Signatures  of a Cold-Flow Disk from MUSE   using a  $z\sim$1 galaxy--quasar pair  towards SDSSJ1422$-$0001~\altaffilmark{*}}
\author{N. Bouch\'e\altaffilmark{1}, 
H. Finley\altaffilmark{2,3}, 
I. Schroetter\altaffilmark{2,3},
M. T. Murphy\altaffilmark{4},
P. Richter\altaffilmark{5,6}, 
R. Bacon\altaffilmark{7},
T. Contini\altaffilmark{2,3}, 
J. Richard\altaffilmark{7}, 
M. Wendt\altaffilmark{5,6}, 
S. Kamann\altaffilmark{8}, 
B. Epinat\altaffilmark{2,9},
S. Cantalupo\altaffilmark{10}, 
L. A. Straka\altaffilmark{11},
J. Schaye\altaffilmark{11},  
C. L. Martin\altaffilmark{12},
C. P\'eroux\altaffilmark{9},
L. Wisotzki\altaffilmark{5}, 
K. Soto\altaffilmark{10},
S. Lilly\altaffilmark{10}, 
C. M. Carollo\altaffilmark{10}, 
J. Brinchmann\altaffilmark{11,13}, 
W. Kollatschny\altaffilmark{8}
}

\altaffiltext{*}{Based on observations made at the ESO telescopes under program 080.A-0364 (SINFONI), 079.A-0600 (UVES) and as part of MUSE commissioning (ESO program 060.A-9100).
Based on observations made with the NASA/ESA Hubble Space Telescope, obtained   at the Space Telescope Science Institute, which is operated by the Association of Universities for Research in Astronomy, Inc., under NASA contract NAS 5-26555. These observations are associated with program ID 12522. }
\altaffiltext{1}{CNRS/IRAP, 9 Avenue Colonel Roche, F-31400 Toulouse, France}
\altaffiltext{2}{CNRS/IRAP, 14 Avenue E. Belin, F-31400 Toulouse, France}
\altaffiltext{3}{University Paul Sabatier of Toulouse/ UPS-OMP/ IRAP, F-31400 Toulouse, France }
\altaffiltext{4}{Centre for Astrophysics and Supercomputing, Swinburne University of Technology, Hawthorn, VIC 3122, Australia}
\altaffiltext{5}{AIP, Leibniz-Institut f\"ur Astrophysik Potsdam, An der Sternwarte 16, 14482 Potsdam, Germany}
\altaffiltext{6}{Institut f\"ur Physik und Astronomie, Universit\"{a}t Potsdam, Karl-Liebknecht-Str. 24/25, 14476 Golm, Germany}
\altaffiltext{7}{Univ. Lyon, Univ. Lyon1, ENS de Lyon, CNRS, Centre de Recherche Astrophysique de Lyon UMR5574, F-69230, 
Saint-Genis-Laval, France}
\altaffiltext{8}{Institut f\"ur Astrophysik, Universit\"at G\"ottingen, Friedrich-Hund-Platz 1, 37077 G\"ottingen, Germany}
\altaffiltext{9}{Aix-Marseille Universit\'e, CNRS, LAM (Laboratoire d’Astrophysique de Marseille) UMR 7326, 13388 Marseille, France}
\altaffiltext{10}{ETH Zurich, Institute of Astronomy, Wolfgang-Pauli-Str. 27, 8093 Zurich, Switzerland}
\altaffiltext{11}{Leiden Observatory, Leiden University, PO Box 9513, 2300 RA Leiden, The Netherlands}
\altaffiltext{12}{Department of Physics, University of California Santa Barbara, Santa Barbara, CA, USA}
\altaffiltext{13}{Instituto de Astrof{\'\i}sica e Ci{\^e}ncias do Espa\c{c}o, Universidade do Porto, CAUP, Rua das Estrelas, PT4150-762 Porto, Portugal}
\begin{abstract}
We use a background quasar to detect the presence of circumgalactic gas  around a $z=0.91$ low-mass star-forming galaxy.
Data from  the new  Multi Unit Spectroscopic Explorer (MUSE)  on the Very Large Telescope  
 show that   the galaxy has a dust-corrected star formation rate (SFR)  of \sfr, with  no companion down to 0.22 \mpy\ (5 $\sigma$) within 240 \kpc\ (30\arcsec).
Using a high-resolution  spectrum (UVES) of the background quasar, which is fortuitously aligned with the galaxy major axis (with an azimuth angle $\alpha$ of only $15^\circ$),
we find, in the gas kinematics traced by low-ionization lines, distinct signatures consistent with those expected for a ``cold-flow disk''
extending  at least 12 kpc ($3\times R_{1/2}$).
We estimate the mass accretion rate $\dot M_{\rm in}$
 to be at least two to three times larger than the SFR, using the geometric constraints from the IFU data and   the \HI\ column density of $\log \NHI/\cmsq \simeq 20.4$ obtained from a {\it Hubble Space Telescope}/COS near-UV spectrum.
From a detailed analysis of the low-ionization lines (e.g. \ZnII, \CrII, \TiII, \MnII, \SiII),
the accreting material appears to be  enriched to about 0.4 $Z_\odot$
(albeit with large uncertainties: $\logZ=-0.4~\pm~0.4$), which is comparable to the galaxy metallicity ($12+\log \rm O/H=8.7\pm0.2$)
implying a large recycling fraction from past outflows.
Blueshifted \MgII\ and \FeII\ absorptions in the galaxy spectrum from the MUSE data   reveals the presence of an outflow.
The \MgII\ and \FeII\ doublet ratios indicate emission infilling due to scattering processes, but the MUSE data do not show any signs of fluorescent \FeII* emission.
\end{abstract}


\section{Introduction}
 A number   of indirect arguments imply that galaxies are  fed by the  accretion of intergalactic gas throughout their evolution.  
For instance, the amount of cold gas present in local and distant galaxies is barely enough to sustain their star formation rates (SFRs) for another Gyr or so 
\citep[e.g.][]{LeroyA_08a,FreundlichJ_13a,TacconiL_13a,SaintongeA_13a}. Another indirect argument comes from the metallicity distribution of G-stars in the Milky Way, which is not consistent with what one finds with `closed-box' chemical evolution models unless some fresh gas infall is invoked 
\citep{Lynden-BellD_75a,PagelB_75a}.  This is often referred to as the G-dwarf problem  \citep{vandenberghS_62a,SchmidtM_63a}.
The very mild evolution of the cosmic neutral density $\Omega_{\HI}$ for damped \Ly\ absorbers  \citep[e.g.][and references therein]{PerouxC_03a,NoterdaemeP_12b,CrightonN_15a}
  together with the rapid evolution of the stellar 
cosmic density, is another indirect argument for continuous replenishment of galaxy reservoirs.


In numerical simulations, accretion of intergalactic gas (via the cosmic web) originates from the growth of dark matter halos which pulls the cold baryons along.  In galaxies with luminosities less than $L^*$, this process is expected to be very efficient 
owing to the short cooling times in these halos \citep{WhiteS_91a,BirnboimY_03a}. 
 This process is expected to lead to distinct  signatures in absorption systems
with  $N_{\HI}$ of $10^{17}$ to $10^{21}$ cm$^{-2}$ 
seen in background quasar sightlines  
\citep{DekelA_09a,KimmT_11a,FumagalliM_11a,FumagalliM_14a,StewartK_11a,GoerdtT_12a,vandeVoortF_12a}.
Once inside the galaxy dark matter halo, the accreted gas is expected to orbit  the galaxy,  delivering not just fuel for star formation
 but  also   angular momentum \citep{StewartK_11a,DanovichM_15a}. 
  In this context, the accreting material coming from the large-scale filamentary structure
should co-rotate with the central disk, forming a warped, extended  gaseous structure
 \citep{PichonC_11a,KimmT_11b,DanovichM_12a,DanovichM_15a,ShenS_13a}, sometimes referred to as a
 ``cold-flow disk'' \citep{StewartK_11a,StewartK_13a}.
In the local universe, such large gaseous disks are often seen around galaxies in \HI\ 21cm surveys,
where the \HI\ disk extends 2--3 times beyond the stellar radius as in  the M33 low surface brightness disk \citep{PutmanM_09a}, and the more massive  M81 \citep{YunM_94a} and M83 galaxies
\citep{HuchtmeierW_81a, BigielF_10a}, among others. 
The kinematics of this \HI\ gas in the outer parts show that it is systematically
rotating in the same direction as the central object.

These gaseous structures ought to  produce distinct kinematic signatures in absorption systems, as argued by \citet{StewartK_11a, StewartK_13a}. 
The infalling gas kinematics is expected to be offset  from the galaxy's systemic velocity when observed in absorption along
background quasar  sightlines \citep{StewartK_11a} because the gas is not rotationally supported. 
These expected signatures are testable against observations with suitably located background sources
such as background quasars \citep{BoucheN_13a} or background galaxies \citep{DiamondStanicA_15a}.

 \citet{BoucheN_13a} presented a first comparison of such inflow kinematics in a $z\simeq2.3$ galaxy--quasar pair 
toward the quasar HE 2243$-$60. The apparent location of this background quasar (and the one presented in this study) is fortuitously
aligned with the galaxy major axis. This configuration is the most favorable situation to look for such inflow kinematic signatures 
since it removes deprojection ambiguities and the geometry allows to rule out any outflow interpretation.
 The data presented in 
 \citet{BoucheN_13a} showed observational signatures similar to theoretical predictions \citep{StewartK_11a,ShenS_13a,DanovichM_15a}.
If the accreting material coming from the large-scale filamentary structure
forms a roughly co-planar structure   around the galaxy with an azimuthal symmetry,
one can infer the amount of gas involved in the process and hence the accretion rate.
The $z\simeq2.3$ galaxy in \citet{BoucheN_13a} was found to have an SFR of $\sim$30~\mpy\ and an accretion rate of 30--60 \mpy.

Other kinematic evidence of gas inflows from red-shifted absorption lines in 
galaxy spectra has been reported by \citet{MartinC_12a} and \citet{RubinK_12a}; however,
these studies lack the critical information on the spatial location of the infalling material with respect to the host.
The recent IFU observations of  \citet{MartinC_15a} of a giant \lya-emitting filament 
around a high-redshift quasar \citep{CantalupoS_14a} provide possible evidence for
kinematics compatible with a large (220 kpc in radius) gaseous rotating disk. 

In this paper, we use a
 quasar--galaxy pair toward the quasar SDSS J142253.31$-$000149 (hereafter SDSS J1422$-$00)
to search for the kinematic signatures of gas inflows.
This quasar is selected from our   SINFONI \MgII\ Program for Line Emitters (SIMPLE)
survey  \citep[][hereafter Paper~I]{BoucheN_07a}.
The SIMPLE survey consists of a search for galaxies around strong $z\sim0.8$--1.0 \MgII\ absorbers
selected from the Sloan Digital Sky Survey (SDSS) database with rest-frame equivalent widths
 $\EW>$2~\AA\ using the IFU SINFONI.
In \citet[][hereafter Paper~II]{SchroetterI_15a}, we analyzed the quasar  apparent location with respect to the host kinematic axis using our \galpak\ algorithm
\citep{BoucheN_15a} and found that this quasar is also fortuitously aligned
with the host galaxy's major axis, as in the $z\sim2$ pair discussed in
  \citet{BoucheN_13a}, at an impact parameter of 12~kpc (1\farcs4) from the host.
This fortuitous alignment makes this quasar--galaxy pair an excellent candidate to study the properties of cold-flow disks.

In order to test the capabilities of the  new  Multi Unit Spectroscopic Explorer (MUSE) instrument \citep{BaconR_06a,BaconR_10a,BaconR_15a} on the Very Large Telescope (VLT), this field was observed during the second commissioning run on 2014 May 6.  
These  observations, covering  \OII\ and \Hb, complement the \Ha$+$\NII\ observations of SINFONI, allowing us to constrain  the 
interstellar medium (ISM) metallicity.
We also have a deep high-resolution VLT/UVES (Ultraviolet and Visual Echelle Spectrograph) spectrum of the background
quasar  and   a near-UV (NUV) spectrum obtained with the G230L grating of the Cosmic Origin Spectrograph (COS)
 on board the {\it Hubble Space Telescope} (HST), allowing us to constrain 
 the metallicity of the absorbing material. Furthermore, 
the UVES kinematics yield   insights into the physical nature of the gas and   show  similar
features to those  in the hydrosimulations of \citet{StewartK_11a} and \citet{ShenS_13a}.

In Section \S~\ref{section:obs}, we present the observations obtained with the  VLT/MUSE instrument (\S~\ref{section:muse})
and the  {\it HST}/COS spectra (\S~\ref{section:cos}).
In Section~\S\ref{section:data}, we present ancillary
SINFONI and UVES  data obtained on this quasar--galaxy pair.
In Section~\S\ref{section:results}, we present the analysis of the IFU data at hand (MUSE and SINFONI), with respect to the host galaxy,
namely, its SFR and  its emission kinematics.
In Section~\S\ref{section:cgm}, we present the analysis of the properties of the circumgalactic gas.
Throughout this paper,
we use  the standard $\Lambda$CDM cosmology with the parameters
$\Omega_{\rm m}=0.3$, $\Omega_{\Lambda}=0.7$, and a Hubble constant $H_0=100\;h$~\kms~Mpc$^{-1}$
with $h=0.7$.

\begin{table}
\caption{Observations Summary\label{table:obs}}

\centering
\begin{tabular}{lcrcl}
 Instrument  & Setting & $T_{\rm exp}$ & PSF & Date of Observation \\
\hline
VLT/UVES	&	390+564		&9000s  & 	1\farcs0		&2007 Apr. 12, 14  \\
VLT/SINFONI &	 J250  		&9600s  &		0\farcs8 	& 2008 Feb 15, 25\\
& & & &  2008 Mar. 14\\
VLT/MUSE  & 	WFM-NOAO-N   	& 7200s & 	0\farcs6		& 2014 May 06 \\
{\it HST}/COS & G230L 		& 11290s & 	---		&2013 Jan. 25\\
\hline
\end{tabular}\par
\bigskip
{ For each instrument we show the setting used,
the exposure time, the point-spread function (PSF) FWHM, and the dates of the observations.}
\end{table}

\section{New Observations}
\label{section:obs}

\subsection{VLT/MUSE}
\label{section:muse}

\begin{figure*}
\centering
\includegraphics[width=16cm,height=14cm]{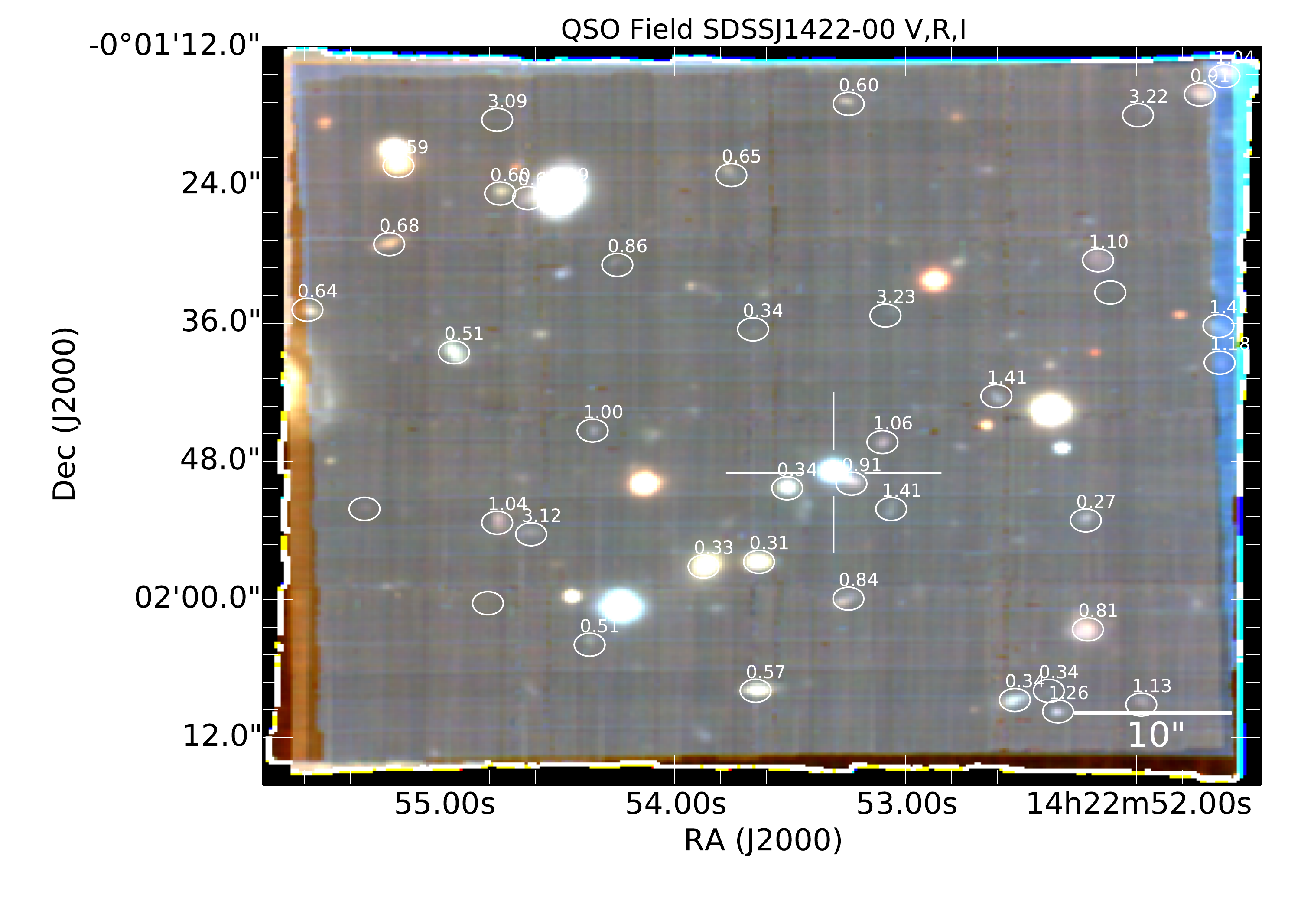}
\caption{Color image made from three broadband images ($V$, $R$ and $I$) extracted from the MUSE data cube 
showing the entire 1\arcmin x1\arcmin\  field of view.
The QSO location is indicated by the cross.
Galaxies with secure redshifts are labeled with white circles. Circles with no label indicate that an emission line is present, but the redshift identification is not secure. Table~\ref{table:catalog} lists all of the sources found in the field.
 }
\label{fig:musefield}
\end{figure*}

\begin{figure*}
\centering
\includegraphics[width=18cm,height=14cm]{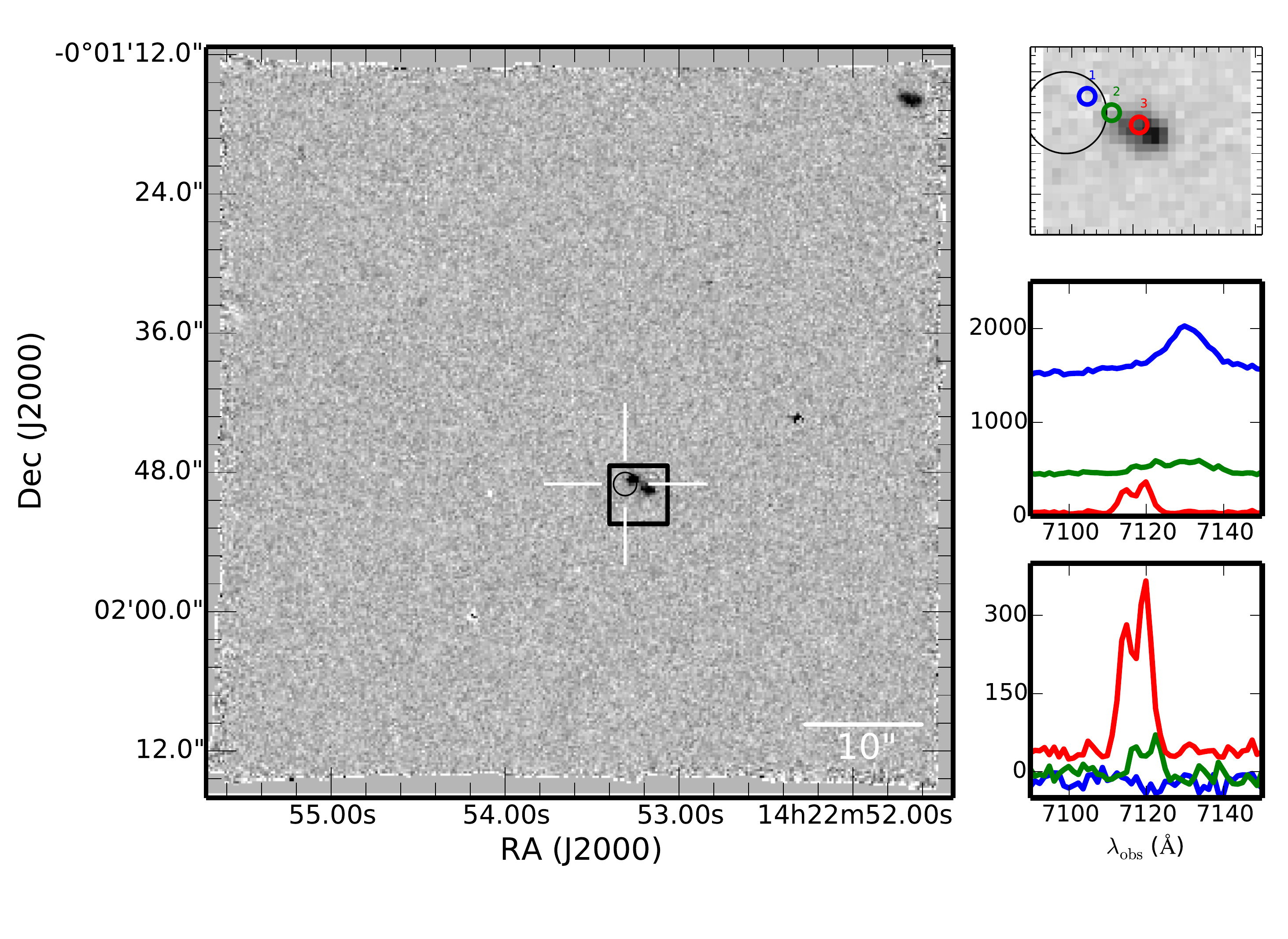}
\caption{Pseudo-narrowband image (7110---7120~\AA)  of the MUSE field made at the expected wavelength of \OII\ 
for the redshift ($z\sim\redshift$)  of the strong \MgII\ absorber with $\EW=3.2$~\AA.
The quasar location is marked by the black circle, which has a radius of 1\arcsec.
There are two \OII\ emitters, one at an impact parameter of 12~kpc (1\farcs4) and another at an impact parameter
of 315~kpc (40\farcs5) in the NW corner.  
The faint feature to the west of the quasar is merely residuals from the continuum subtraction of a star.
 The 5\arcsec x5\arcsec\ black square box represents the subfield that will be used in Figure~\ref{fig:maps} and  in the top inset. 
The top subpanel shows a narrowband image (7110--7120~\AA) with the QSO PSF removed (see text for the details). 
The middle subpanel shows the spectra around \OII\ at the three labeled positions prior to the QSO PSF subtraction, 
revealing the \NeV$\lambda 3425$ emission from the QSO.
The bottom subpanel shows the  spectra (in units of $10^{-20}$ \fdlam) around \OII\ at the three labeled positions after the QSO PSF subtraction.
The three spectra (blue, green, and red) are from the three positions labeled 1--3 in the top inset, corresponding to the  
quasar location (``1''), the overlap region (``2'') and the center of the galaxy (``3'').
 }
\label{fig:musefield:narrow}
\end{figure*}

\begin{table}
\caption{MUSE Sources in the SDSSJ1422$-$00 field }
\label{table:catalog}
\centering
\begin{tabular}{ccccc}
ID & R.A. & Decl. & Redshift & Lines \\
\hline
obj029 & 14:22:51.619 & -00:01:14.53 & $\cdots$ & 7619 \\
obj042 & 14:22:51.640 & -00:01:39.43 & 1.176 & \OII \\
obj035 & 14:22:51.645 & -00:01:36.24 & 1.406 & \OII \\
obj028 & 14:22:51.726 & -00:01:16.13 & 0.909 & \OII \\
obj033 & 14:22:51.979 & -00:02:09.13 & 1.130 & \OII \\
obj006 & 14:22:51.993 & -00:01:17.93 & 3.224 & \Ly \\
obj008 & 14:22:52.113 & -00:01:33.33 & $\cdots$ & 5160. \\
obj032 & 14:22:52.166 & -00:01:30.53 & 1.100 & \OII \\
obj039 & 14:22:52.219 & -00:01:53.13 & 0.266 & \OIII,\Hb \\
obj023 & 14:22:52.233 & -00:02:02.33 & 0.811 & \OII,\OIII,\Hb \\
obj034 & 14:22:52.339 & -00:02:09.73 & 1.259 & \OII \\
obj003 & 14:22:52.379 & -00:02:07.94 & 0.345 & \Ha,\Hb,\OIII,\OII \\
obj004 & 14:22:52.526 & -00:02:08.73 & 0.345 & \Ha,\Hb,\OIII,\OII \\
obj036 & 14:22:52.606 & -00:01:42.33 & 1.405 & \OII \\
obj037 & 14:22:53.061 & -00:01:52.14 & 1.407 & \OII \\
obj007 & 14:22:53.086 & -00:01:35.33 & 3.227 & \Ly \\
obj031 & 14:22:53.099 & -00:01:46.33 & 1.059 & \OII \\
obj027 & 14:22:53.233 & -00:01:49.93 & 0.909 & \OII \\
obj014 & 14:22:53.245 & -00:01:16.94 & 0.600 & \OII,\OIII,\Hb \\
obj024 & 14:22:53.246 & -00:01:59.93 & 0.839 & \OII,\OIII,\Hb \\
obj002 & 14:22:53.511 & -00:01:50.34 & 0.345 & \Ha,\NII,\OIII,\Hb,\OII \\
obj019 & 14:22:53.633 & -00:01:56.73 & 0.309 & CaHK \\
obj011 & 14:22:53.659 & -00:02:07.93 & 0.575 & \OII,\OIII,\Hb,Hg... \\
obj022 & 14:22:53.659 & -00:01:36.53 & 0.340 & \OIII,\Hb, \\
obj015 & 14:22:53.753 & -00:01:23.13 & 0.652 & \OII,\OIII \\
obj018 & 14:22:53.873 & -00:01:57.14 & 0.334 & CaHK \\
obj025 & 14:22:54.246 & -00:01:30.93 & 0.859 & \OII \\
obj038 & 14:22:54.353 & -00:01:45.33 & 0.996 & \OII \\
obj026 & 14:22:54.366 & -00:02:03.93 & 0.512 & \OII,\OIII,\Hb \\
obj001 & 14:22:54.499 & -00:01:24.74 & 0.188 & \Ha,\SII,\NII,\OIII \\
obj005 & 14:22:54.619 & -00:01:54.33 & 3.122 & \Ly \\
obj021 & 14:22:54.633 & -00:01:25.13 & 0.676 & \OII,\OIII,\Hb \\
obj012 & 14:22:54.753 & -00:01:24.73 & 0.600 & \OII,\OIII,\Hb \\
obj016 & 14:22:54.766 & -00:01:18.33 & 3.086 & \Ly \\
obj030 & 14:22:54.766 & -00:01:53.33 & 1.043 & \OII \\
obj009 & 14:22:54.806 & -00:02:00.33 & $\cdots$ & 5528. \\
obj010 & 14:22:54.953 & -00:01:38.53 & 0.513 & \OII,\OIII,\Hb,Hg... \\
obj017 & 14:22:55.193 & -00:01:22.33 & 0.588 & CaHK \\
obj020 & 14:22:55.233 & -00:01:29.13 & 0.675 & \OII,CaHK \\
obj040 & 14:22:55.587 & -00:01:34.83 & 0.641 & \OII,\OIII \\
obj041 & 14:22:55.340 & -00:01:52.13 & $\cdots$ & 8438 \\
\hline
\end{tabular}\\
{For each source, we list the J2000 coordinates, the redshift, and the line(s) used in securing the redshift.
When the line is unique or its shape ambiguous, we list the observed wavelength in \AA.}
\end{table}

This  $z=1.08$  quasar
was observed with the new wide-field (1\arcmin~x~1\arcmin) IFU for the VLTs
\citep[MUSE,][]{BaconR_10a}
 during the second  commissioning run  on 2014 May 6 under good seeing conditions (FWHM$\sim$0\farcs6)
for 2~hr, in $4\times30$~min exposures (Table~\ref{table:obs}).
The data were reduced with the MUSE pipeline~\footnote{A short description of the pipeline is given in \citet{WeilbacherP_12a}.} 
 v1.0 with  standard settings. 
 We used the  bias, flat-field calibrations, and arc lamp exposures taken during the day for that night.
The wavelength solution is calibrated on the air scale. To minimize flat-field errors from spatial shifts related to temperature changes
during the night, we only use the flat fields  that were taken when the  temperature was within $\pm0.5^{\circ}$ C from the ambient temperature of the observations (mean of 15.1$^{\circ}$ C). We used four twilight flats, each rotated at 90$^\circ$,  and corrected for vignetting using the vignetting mask. With these calibrations, we processed the raw science data using the MUSE recipes {\it scibasic} and {\it scipost} with the sky removal option turned off to produce data cubes and pixel tables for each of the four exposures.

The individual exposures were registered  using the point sources in the field, ensuring
accurate relative astrometry, as significant shifts of a few tens of arcsecs can occur owing to  the  spatial shifts introduced by the derotator wobble between exposures.
 The pixel tables from the individual exposures were then combined
to a single data cube produced  using a 3-dimensional (3D)
drizzle interpolation process. The MUSE data cube is sampled to a common grid (0\farcs2$\times$0\farcs2$\times1.25$~\AA)
and the final wavelength solution is calibrated on air and   corrected for the heliocentric velocity.
The cube is available at this URL: \url{http://muse-vlt.eu/science/j1422}.

We checked the wavelength solution against the wavelength of OH lines and found it to be accurate {within 10 \kms}.
We forced the final astrometry solution to match the SDSS coordinates using the
point sources in the field. 
The sky subtraction was performed on the combined exposure with the Zurich Atmosphere Purge (ZAP) principal component algorithm developed by \citet{SotoK_16a},
which was designed  to remove  OH line  residuals.
The flux calibration was obtained from observations of the spectrophotometric standard star 
GD~108. The night was photometric, and we cross-checked the flux measurements against the SDSS
magnitudes by fitting a Moffat function to the  stars in the field in reconstructed images obtained with the SDSS filter curves. 
We found no difference between the SDSS magnitudes and our measurements greater than 0.01 magnitude.

From the noise in the data at the expected wavelength of \OII\ ($\sim$7118~\AA),  $2.3\times10^{-20}$~\fdlam\ (1$\sigma$) per pixel, 
we estimate 
a   surface brightness limit of $1.5\times10^{-18}$~\flux~arcsec$^{-2}$ (1$\sigma$) for  emission-line objects (FWHM = 2.14$\times$1.25~\AA), which corresponds to a flux limit of $6\times10^{-19}$~\flux\ (1$\sigma$) for an unresolved line emitters at 0\farcs7 seeing.
For the \OII\ doublet,  the flux limit is twice this value, or $\sim1.2\times10^{-18}$~\flux\ (1$\sigma$).
Hence, our 5$\sigma$ limit for unresolved \OII\ emitters at $z\approx0.91$ corresponds to a SFR of 0.22 \mpy\ using the \citet{KewleyL_04a} calibration
with no dust-reddening (as in Section 4.3).
The flux limit for an unresolved continuum emission ($3\times3$ spaxel) is  $\sim8\times10^{-20}$~\fdlam\ (1$\sigma$) 
corresponding to  $\sim$25.8 (24.7; 3$\sigma$) AB-magnitudes at 7100~\AA. 
Note  that the MUSE sensitivity is weakly dependent on  wavelength, outside regions affected by sky emission lines.

Figure~\ref{fig:musefield} shows the MUSE data, with
a color image  made from three broadband images ($V$, $R$, and $I$) extracted from the   data cube.  
The quasar location is represented by the cross. 
For completeness, we have searched for all  galaxies using both a visual inspection of the cube and a SExtractor-based algorithm (MUSE Line Emission Tracker [MUSELET], Richard et al., in prep.),
and we  found  41 galaxies with emission or absorption lines, of which 37 have a reliable redshift. The coordinates and redshifts of these galaxies are listed in Table~\ref{table:catalog},
and their locations are shown in Figure~\ref{fig:musefield} with the redshifts labeled.

Here  we are only interested in   galaxies with redshifts comparable to the \MgII\ absorption redshifts at $z\approx0.91$, and
the MUSE field of view (1\arcmin$\times$1\arcmin) allows us to investigate whether the \MgII\ absorption could be associated
with other host galaxies that would have fallen outside the SINFONI field of view (8\arcsec$\times$8\arcsec).  Figure~\ref{fig:musefield:narrow} shows a pseudo-narrowband image centered on the expected \OII\ wavelength of 7110--7120~\AA\ (8 spectral pixels)  with a linear continuum subtraction from  the MUSE data, where the quasar SDSS~J1422$-$00 is marked by the cross 
and black circle.
Only two \OII\ emitters are detected,  one at
an impact parameter of only $b=$1\farcs45 ($\sim$12~kpc) and 
another at 40\farcs5 ($\sim 315$ kpc) in the NW corner of the field. 
The second one is at a  large distance (about $3.3\times R_{\rm vir}$), implying that it is likely unrelated to 
the absorbing gas.
There are no other \OII\ emitters within $\pm5000$~km/s down to 0.22 \mpy\ (5$\sigma$),
 and   the objects visible in white correspond to
imperfect continuum subtraction of stellar objects with strong continuum slopes.

\subsection{{\it HST}/COS}
\label{section:cos}

In order to characterize the \HI\ gas column density probed by the background quasar, we obtained 
a   spectrum of the quasar with the NUV G230L grating of the {\it HST}/COS instrument 
 (Cycle 19, program ID 12522,  PI: N. Bouch\'e) covering the \Ly$~\lambda$1216 transition   at $\lambda_{\rm obs}=2321$~\AA.
The {\it HST}/COS spectrum was  obtained on   2013 January with four orbits  for a total exposure time of 
$T_{\rm exp}=11,290$~s (Table~\ref{table:obs}).
The spectrum has a resolution of $R\sim2900$ In the wavelength range 2100--2550~\AA\ of stripe A of the G230L grating.
%
%
The final spectrum was reduced with the COS  {\it calcos} \citep{KaiserM_08a} pipeline
and  has a signal-to-noise ratio (SNR) of {$\sim3$}
  around the wavelength of the redshifted \Ly\  absorption line.

\section{Ancillary Data }
\label{section:data}
 
\subsection{VLT/SINFONI}

The original  VLT/SINFONI  data covering the quasar  J1422$-$00 (Paper~I) 
  revealed the host galaxy at
$z=\redshift$ in shallow (40-minute integration time) exposures.
We reobserved the field with SINFONI for 2.6hr  in 2008 (Table~\ref{table:obs}).
In paper~II, we presented 
the analysis of these  VLT/SINFONI data of this field along with the   13 other quasar fields making the SIMPLE sample.
The data reduction was performed  as in paper~I and \citet{ForsterSchreiberN_09a},
 using the  SINFONI pipeline \citep[SPRED,][]{SchreiberJ_04a,AbuterR_06a} and
 complemented with  custom routines for OH sky-line removal  \citep{DaviesR_06a} and using the Laplacian edge cosmic-ray removal technique of \citet{vanDokkumP_01a}.
The SINFONI data cube is sampled at  0\farcs125$\times$0\farcs125$\times1.4$~\AA\ and the wavelengths are calibrated on the vacuum scale.

\subsection{VLT/UVES}

As part of the SIMPLE survey, the quasar   was observed om 2007 (Table~\ref{table:obs}) with the high-resolution VLT/UVES spectrograph. 
The VLT/UVES data were taken with the 390$+$564 nm central wavelength setting. The data were reduced using
 version 3.4.5 of the UVES pipeline in MIDAS, and the data reduction details were presented 
in Paper~II.

\section{Host galaxy Properties}
\label{section:results}

\subsection{QSO PSF Subtraction}
\label{section:qsopsf}

Given the small impact parameter (1\farcs4), the quasar continuum emission overlaps spatially 
with the \MgII-absorber galaxy emission (see Figure~\ref{fig:musefield:narrow}).
 We therefore need to carefully remove the quasar continuum in the MUSE data before performing any detailed kinematic analysis.
The continuum subtraction task is complicated by an  \NeV$\lambda3425$ emission from the QSO,
which appears at around the same wavelength (7115~\AA)
 as the \OII$\lambda\lambda3727,3729$ doublet from the host galaxy.

To remove the QSO continuum that overlaps with the \OII\ emitter,  we constructed a 3D PSF  using the PampelMuse algorithm 
\citep{KamannS_13a} to  interactively and simultaneously fit  the QSO continuum and the PSF. 
We then removed this 3D PSF from the cube.
The resulting narrowband image is shown in the top right subpabel of Figure~\ref{fig:musefield:narrow}.
The other two subpanels in Figure~\ref{fig:musefield:narrow} show three spectra  taken at the three positions labeled in the first inset
before (middle) and after (bottom)  the 3D PSF subtraction.
These spectra show that the \OII\ emission in the overlap region (``2'') becomes clearly apparent  after the PSF subtraction.

\subsection{Host galaxy redshift}

Much of the analysis presented in the following sections  depends on the redshift of the host  and its accuracy.
Therefore,  we use several methods to cross-check our measurements and remove possible systematic errors.
As in paper~II, we determine the redshift from the mean  wavelength of
the reddest and the bluest parts of the \OII\ emission  (along the kinematic major axis) and from 
a pseudo-long slit aligned with the kinematic major axis. 
We also estimated the redshift from \Ha\ in the SINFONI data using a similar technique and found
 $z\simeq0.9096$ (Paper~II). 

\begin{figure}
\centering
\includegraphics[width=8cm]{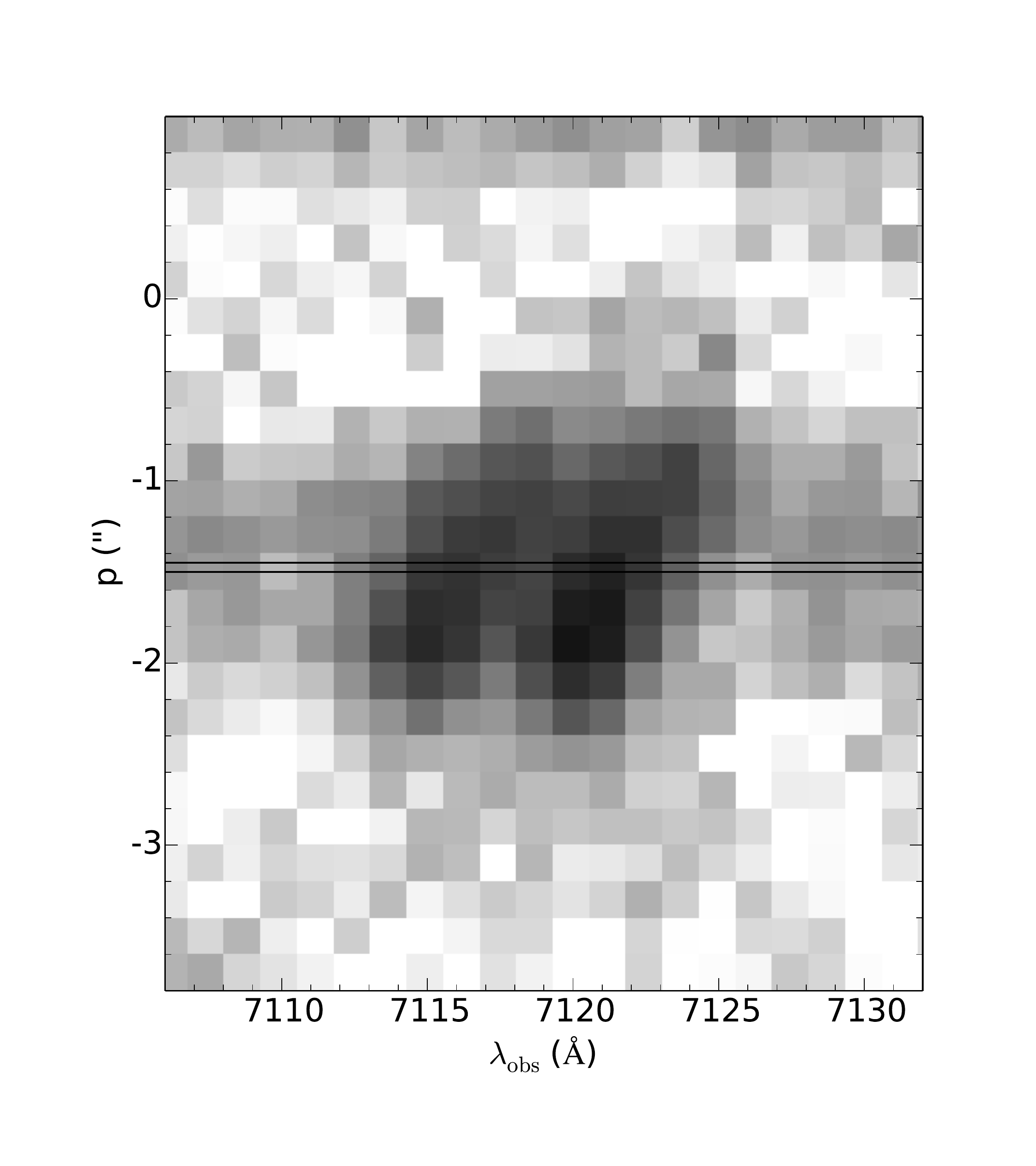}
\caption{Position-velocity ($p-v$) diagram extracted from the MUSE data
along a pseudo-longslit spectrum around the \OII\ emission line covering both the QSO and galaxy location (PA$\;=55^\circ$)
where the QSO PSF was subtracted as described in \S~\ref{section:qsopsf}.
The QSO trace position is at $y=0$, and the galaxy trace is seen at an impact parameter of $\sim1\farcs4$ (12 kpc).
For the reddest transition of the \OII\ doublet,
at $\lambda_{\rm rest}=3728.8$~\AA,
the galaxy  systemic velocity is found at $\sim$7120.5~\AA, corresponding to $z=\redshift$.  \label{fig:pv}
}
\end{figure}

Figure~\ref{fig:pv} shows the pseudo-long slit 2-dimensional
 spectra extracted from the MUSE QSO-subtracted data cube
 with a 1\arcsec\ slit passing through the location of the QSO and the galaxy. 
The continuum trace of the galaxy is visible at an impact parameter of $\sim 1\farcs45$ (12 kpc) and is marked by the horizontal line.
The best redshift from the \OII\ MUSE data is estimated from the reddest \OII\ component ($\lambda3728.8$~\AA),
 whose kinematic center appears to be at 7120.5~\AA, corresponding to   \redshift.
The global kinematic fit discussed in the next section yields a redshift consistent with this value.

Hence, we adopt a systemic redshift of the \MgII\ host galaxy of \redshift.

\subsection{Fluxes, SFR }

\begin{figure}
\centering
\includegraphics[width=9cm,height=10.5cm]{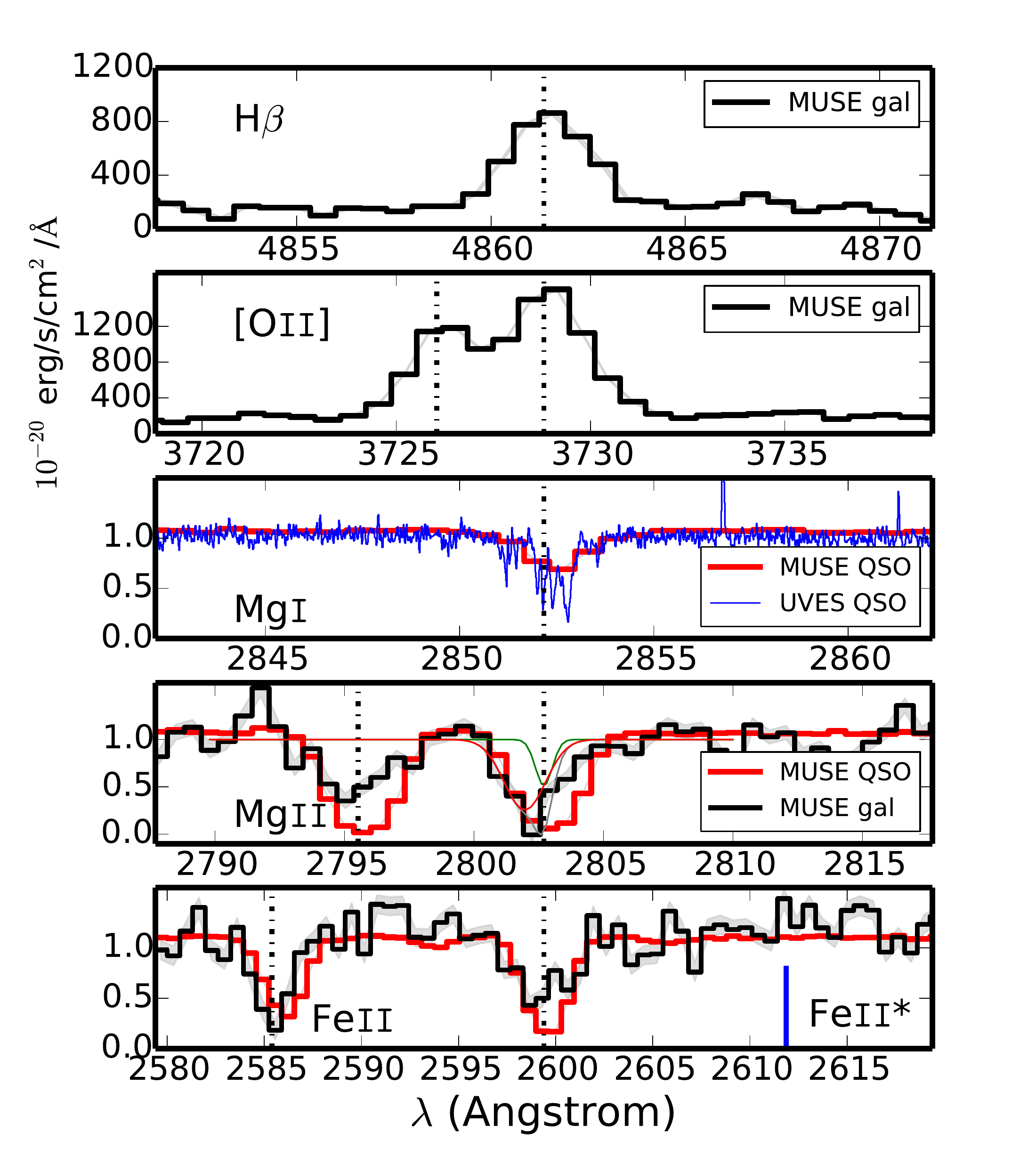}
\caption{MUSE spectrum. 
The black solid lines show the galaxy spectrum extracted from the MUSE data cube in a circular
 aperture of 0\farcs6 in radius.  The red solid lines show the background quasar spectra extracted from the MUSE data cube in a circular  aperture 
with radius 0\farcs6. The panels show the \Hb\ and \OII\ emission and the \MgI, \MgII, and \FeII\ absorption, from top to bottom, respectively.
In all panels,  the systemic redshift (\redshift; see text) is represented by the vertical dotted line. 
The UVES quasar spectrum for \MgI\ is shown for comparison in the middle \MgI\ panel, after converting the UVES spectra to air wavelengths.
The \MgI\ and \MgII\ absorptions in the background quasar are redshifted with respect to the galaxy systemic velocity.
A two-component fit (see \S~\ref{section:wind}) reveals that the \MgII\ absorption in the galaxy spectrum  (second panel from bottom) 
 is blueshifted by $v_{\rm out}\simeq -80\pm15$~\kms\ with respect to the systemic velocity shown by the vertical dotted line,
with a tail up to $-150$ \kms.   
The \FeII$\lambda\lambda2587,2600$ absorption  (bottom panel) is consistent with the \MgII\ profile. 
The line ratios of the \FeII\ and \MgII\ doublets support the presence of infilling emission, but the spectrum
does not reveal any signs for fluorescent \FeII* emission as discussed in \S~\ref{section:wind}.
\label{fig:muse:spectra}}
\end{figure}

Our  VLT/SINFONI  data toward the $z_{\rm qso}=1.083$ quasar  SDSS J1422$-$00 already revealed the host galaxy of the strong \MgII\ absorber with $\EW=3.2$~\AA\ at
$z\simeq0.9096$ (Papers~I, II), whose \Ha\ flux of $9.0\pm0.1\cdot10^{-17}$ \flux\ corresponds to 
an observed SFR$_{\Ha}$ of 2.8$\pm0.2$ \mpy\
 assuming a Salpeter initial mass function (IMF) from 0.1 to 100 \msun\ and applying no dust correction.
With an extinction of $E(B-V)=0.1$ (see below), the intrinsic SFR is found to be 3.5$\pm$0.2 \mpy.
The SINFONI data did not reach the sensitivity required to detect the \NII\ emission, but our data did allow us to obtain a $2\sigma$ upper limit of $f_{\NII}<3\times10^{-17}$~\flux. 

From the VLT/MUSE data, we find (top two panels in Figure~\ref{fig:muse:spectra})
that the \OII\ flux is $1.1\pm0.2\times10^{-17}$ \flux\ and the \Hb\ flux is
$3.3\pm0.3\times10^{-17}$ \flux. 
We do not detect \OIII$ \lambda$4363, which leads to a  $2\sigma$ upper limit
of $<1\times10^{-18}$ \flux.
These total flux measurements (summarized in Table~\ref{table:fluxes}) were obtained from the global 3D line fitting
to the MUSE data described in the next section (\S \ref{section:kinematics}). We also used the traditional `growth curve'    technique
 to verify these values.
The \OII\ luminosity corresponds to a SFR$_{\OII}$ of 3.0$\pm0.1$ \mpy\ using the revised calibration of  \citet{KewleyL_04a}
which makes no assumption about reddening. 

 Taking the \Ha\ and \Hb\ fluxes at face value, i.e. ignoring 
possible systematics in the flux calibration between MUSE (accurate to 0.01 mag) and SINFONI (accurate to 0.15~mag or 15\%), the extinction is $E(B-V)=0.1\pm0.1$ from the Balmer decrement.
Using the Balmer decrement for the reddening, the extinction at \OII\ is 0.5~mag, yielding an intrinsic SFR (SFR$_0$) of around \sfr\ for a Salpeter IMF ranging from 0.1 to 100 \msun
 ~\footnote{The SFR$_0$ would be $\approx5.0$ \mpy\ using the original \citet{KennicuttR_98a} calibration, which includes a dust correction. }.
The flux ratio \OII/\Ha\ ($\approx1$) and our low dust estimate are  entirely consistent with the observations of local galaxies
from \citet{SobralD_12a} and  \citet{KewleyL_04a},
which showed that the  \OII/\Ha\ ratio is strongly dependent on the Balmer decrement,
with  \OII/\Ha\ around $\sim1$ where the Balmer decrement  \Ha/\Hb\ is $\sim3$,
as in our data.

Hence, the SFR estimates from \Ha\ and \OII\ are consistent with each other, and we adopt an SFR of  \sfr\
for a  \citet{ChabrierG_03a} IMF. Table~\ref{table:galaxy} summarizes the extinction and SFR measurements.
Using the galaxy half-light radius  of $R_{1/2}=4.0\pm0.2$ kpc found in \S~\ref{section:kinematics}, the SFR surface density is 
$\Sigma_{\rm SFR}\simeq 0.05\pm0.02$ \mpy~kpc$^{-2}$, where the uncertainty is dominated by
the SFR uncertainties.

\begin{table}
\caption{Host Galaxy Emission and Absorption  Lines   \label{table:fluxes}}
\centering
\begin{tabular}{lcccc}
\hline
& Flux   &  \multicolumn{2}{c}{Instrument} \\
&( \flux) & & \\
\hline
$f_{\Ha,6564}$  &  ($9.0\pm0.1)\;\times10^{-17}$  & \multicolumn{2}{c}{VLT/SINFONI}  \\
$f_{\NII,6583}$  &  $<3\times10^{-17}$ ($2~\sigma$) & \multicolumn{2}{c}{VLT/SINFONI} \\
$f_{\OII,3727}$  &   ($1.1\pm0.2)\times10^{-16}$ & \multicolumn{2}{c}{VLT/MUSE} \\
$f_{\Hb,4861}$   &  ($3.3\pm0.3)\times10^{-17}$ & \multicolumn{2}{c}{VLT/MUSE} \\
$f_{\OIII,4363}$    &    $<1\times 10^{-18}$ ($2~\sigma$)& \multicolumn{2}{c}{VLT/MUSE}\\
\hline
 & $W_r$ & \\
 & (\AA) & \\
\hline
\MgII\ $\lambda{2796}$  & 3.5 $\pm$ 0.4  & \multicolumn{2}{c}{VLT/MUSE}\\
\MgII\ $\lambda{2803}$  & 3.7 $\pm$ 0.4  & \multicolumn{2}{c}{VLT/MUSE}\\
\MgII\ $\lambda{2587}$ & 2.5 $\pm$ 0.4 & \multicolumn{2}{c}{VLT/MUSE} \\
\FeII\ $\lambda{2600}$ & 3.9 $\pm$ 0.4 & \multicolumn{2}{c}{VLT/MUSE} \\
\FeII* $\lambda{2612} $ & $<0.8$  ($2~\sigma$)& \multicolumn{2}{c}{VLT/MUSE}\\
\FeII* $\lambda2626 $ & $<0.8$ ($2~\sigma$)& \multicolumn{2}{c}{VLT/MUSE}\\
\FeII* $\lambda2632 $ & $<0.8$ ($2~\sigma$)& \multicolumn{2}{c}{VLT/MUSE}\\
\hline
\end{tabular}
\end{table}

\begin{table}
\centering
\caption{Geometry and Galaxy Derived Properties\label{table:galaxy}}
\begin{tabular}{lcccc} 
\hline
Quasar $b$ (kpc) & 12 (1\farcs4) & \\
gal-qso P.A. ($^\circ$) & 56 $\pm$ 2 \\
gal. P.A.($^\circ$) & 71 $\pm$ 3 \\
$\alpha$ ($^\circ$) & 15 $\pm$ 2\\
gal. incl. ($^\circ$)& 60 $\pm$ 2\\
\hline
$E(B-V)$ &  0.1 $\pm$ 0.1~\footnotemark[1]& \\
SFR$_{\Ha, \rm obs}  (\mpy) $ & 2.8 $\pm$ 0.2~\footnotemark[2] \\
SFR$_{\Ha, 0}  (\mpy) $ & 3.5 $\pm$ 2.0~\footnotemark[3] \\
SFR$_{\OII, \rm obs}  (\mpy) $ & 3.0 $\pm$ 0.2~\footnotemark[2] \\
SFR$_{\OII,0}$ (\mpy) &  4.7 $\pm$ 2.0~\footnotemark[3] \\
SFR$_{\OII,0}$ (\mpy) &  2.5 $\pm$ 1.0~\footnotemark[4]  \\
$\Sigma_{\rm SFR}$ (\msun~kpc$^{-2}$) & 0.05 & \\
\hline
$\log (N/O)$ & $-1.3~\pm$ 0.3~\footnotemark[5] ($<-0.9$)~\footnotemark[6] \\
$12+\log (\rm{O/H})$ & 8.7 $\pm$ 0.2~\footnotemark[7]   \\
\hline
$R_{1/2}$ (kpc) & 4 $\pm$ 0.2& \\
$R_{\rm vir}$ (kpc) & 90 $\pm$ 5 &\\
$V_{\rm max}$ (\kms) & 110 $\pm$ 10 & \\
$M_{\rm dyn}(<R_{1/2})$ ($10^{10}$\msun) & 2 $\pm$ 0.4 & \\
$M_{\rm h}$ ($10^{11}$\msun) & 1.9 $\pm$ 0.5 & \\
$M_{\rm bar}$ ($10^{10}$\msun) & 0.5 $\pm$ 0.1 & \\
$\lambda_{\rm gal}$ & 0.04 & \\
\hline
$V_{\rm wind}$ (\kms) & 100--150 \\
$\dot M_{\rm wind}$ (\mpy) & 0.5--5 & \\
\hline
$R_{\rm in}$ (kpc) & $\geq12$ &\\
$V_{\rm in}$ (\kms) & $\approx$100 \\
$\dot M_{\rm in}$ (\mpy) & $\approx$10  & \\
$\lambda_{\rm cfd}$ & $>0.06$ &\\
\hline
\end{tabular}
\footnotetext[1]{From the \Hb/\Ha\ flux ratio.} 
\footnotetext[2]{For a Salpeter IMF from 0.1 to 100 \msun.}
\footnotetext[3]{For a Salpeter IMF from 0.1 to 100 \msun\ with $E(B-V)=0.1$.}
\footnotetext[4]{For a  \citet{ChabrierG_03a} IMF from 0.1 to 100 \msun\ with $E(B-V)=0.1$.}
\footnotetext[5]{From  \citet{PerezMonteroE_14a}.}
\footnotetext[6]{Using \NII/\OII\ from \citet{PerezMonteroE_09b}.}
\footnotetext[7]{Using \OII/\Hb\ from \citet{MaiolinoR_08a}.}
\end{table}

\begin{table}
\caption{Host Galaxy  Kinematics \label{table:galpak}}
\centering
\begin{tabular}{lcccc}
\hline
 & \Ha & \OII\ & \Hb\\
\hline
Seeing (arcsec) &  0.78  & 0.55 & 0.55 \\
S/N pix$^{-1}$ (max) & 4.8  &  45 & 19 \\
$R_{1/2}$ (kpc) &  4.0 $\pm$ 0.2 	& 3.9 $\pm$ 0.2 		& 2.7 $\pm$ 0.3\\ 
 incl.  ($^\circ$)  	 & 57 $\pm$ 2 		& 61 $\pm$ 2 		& 41 $\pm$ 2 \\
 P.A. ($^\circ$)  	& 80 $\pm$ 2 		& 71 $\pm$ 2   		& 75 $\pm$ 3 \\
 $V_{\rm max}$ (\kms) \footnote{With the turnover radius $r_{\rm t}$ fixed to 1.5 kpc.}
			& 123 $\pm$ 5  	& 100 $\pm$ 10		&  105 $\pm$ 10 \\
$r_{\rm t}$  (kpc)  & 1.5 & 1.5 & 1.5\\
 $\sigma_o$ (\kms) &  42 $\pm$ 5 	& 34 $\pm$ 2 		& 32 $\pm$ 3  \\ 
\hline
\end{tabular}\par
{These morphological and kinematic parameters are determined from our 3D fits using our  \galpak\ algorithm. }
\end{table}

\subsection{Metallicity}
\label{section:metallicity}

\begin{figure}
\centering
\includegraphics[width=8cm]{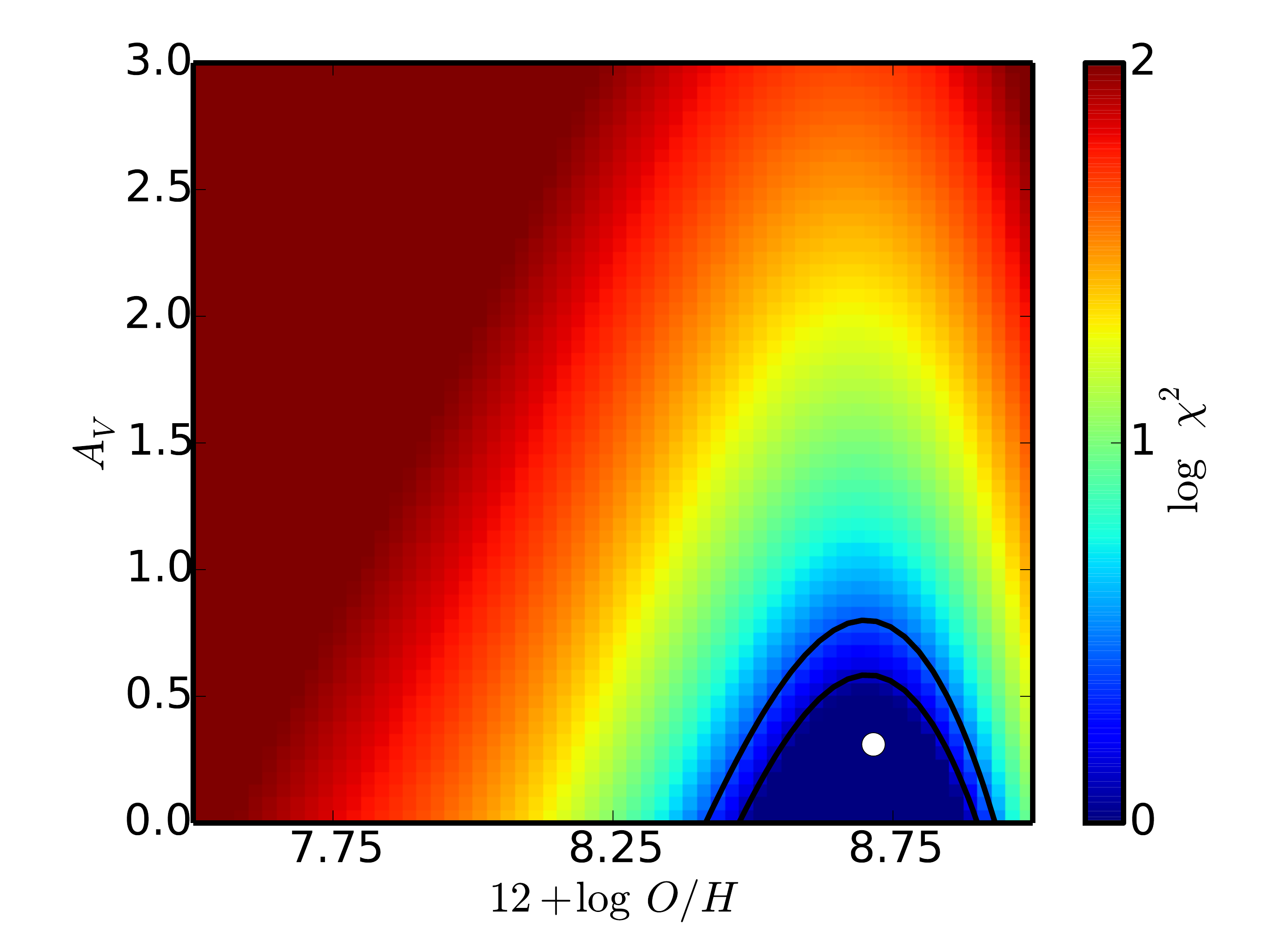}
\caption{Likelihood contours for the metallicity   using the   \NII/\Ha\ and \OII/\Hb\ constraints with $E(B-V)$ set to 0.10 determined from the Balmer decrement.  
The contours show that the metallicity estimate $12+\log \rm O/H=8.7\pm0.2$ is robust against errors in the extinction estimate, because 
   the \OII/\Hb\ ratio is sampling the peak of the relation between  \OII/\Hb\ and metallicity. }\label{fig:metallicity}
\end{figure}

From the nebular line ratios, \NII/\Ha\ and \OII/\Hb\ (Table~\ref{table:fluxes}), we can constrain the galaxy metallicity.
Figure~\ref{fig:metallicity} shows the likelihood contours allowed by the data from the \NII/\Ha\ and \OII/\Hb\
measurements, yielding an extinction   consistent with zero.
This is driven by the \OII/\Hb\ ratio being already above the maximum value between \OII/\Hb\ and metallicity \citep[e.g.][]{MaiolinoR_08a}, and any dust reddening will increase this ratio further, implying that the global fit yields no additional constraint on dust reddening.
In the previous section, we argued that $E(B-V)\simeq0.1$ from the Balmer decrement, which is shown by the white circle in Figure~\ref{fig:metallicity}.
Hence,  the relation  between the  \OII/\Hb\ ratio and metallicity
 imposes a metallicity $12+\log \rm O/H=8.7\pm0.2$, regardless of the extinction value~\footnote{
Our dust-corrected fluxes with the calibration of \citet{PerezMonteroE_14a} yield a
metallicity of $12+\log \rm O/H=8.5\pm0.2$,   consistent within the errors with our value, and $\log(N/O)=-1.3\pm0.3$. 
Using our upper limit on \NII\ and   \OII,
we find an upper limit on the $\log(N/O)$  ratio of $<-0.9$ using the \citet{PerezMonteroE_09b} calibration.  
}, at the peak of    the relation between  \OII/\Hb\ and metallicity.
Hence, the metallicity estimate is robust against the reddening estimate as illustrated by the contours
in Figure~\ref{fig:metallicity}.
Systematic uncertainty remains in the metallicity absolute calibration, as the metallicity might depend on the $N/O$ abundances,
as argued by \citet{PerezMonteroE_09b} and \citet{PerezMonteroE_13a}. 
We conclude that  the  ISM of this galaxy is enriched at a metallicity $\logZ=0.0\pm0.2$  using
the solar value $12+\log \rm O/H=$8.7 for oxygen \citep{AsplundM_09a}.

Spatial variations of the  \OII/\Hb\ flux  ratio might indicate the presence of a metallicity gradient.
However, the weaker \Hb\  line has a smaller S/N than \OII\ and falls at 9280~\AA, close to the end of the wavelength coverage of MUSE, where the MUSE sensitivity drops sharply.
Given the difficulty in mapping the \Hb\ line in this part of the spectrum, we fitted   \OII\ and \Hb\ jointly with the CAMEL
algorithm of  \citet{EpinatB_12a}. 
The resulting \OII/\Hb\ map shows no variations along the galaxy major axis, but shows a possible gradient along the galaxy minor axis:
the   ratio increases from 2.3$\pm0.1$ in the center to  3.0--4.0 at the edges. This could be  due to variations in ionization conditions
or to  systematic errors from the weaker \Hb\ line, which also appears to be more compact (Table~\ref{table:galpak}).

\subsection{Galaxy  Kinematics}
\label{section:kinematics}

\begin{figure*}
\centering 
\includegraphics[width=16cm]{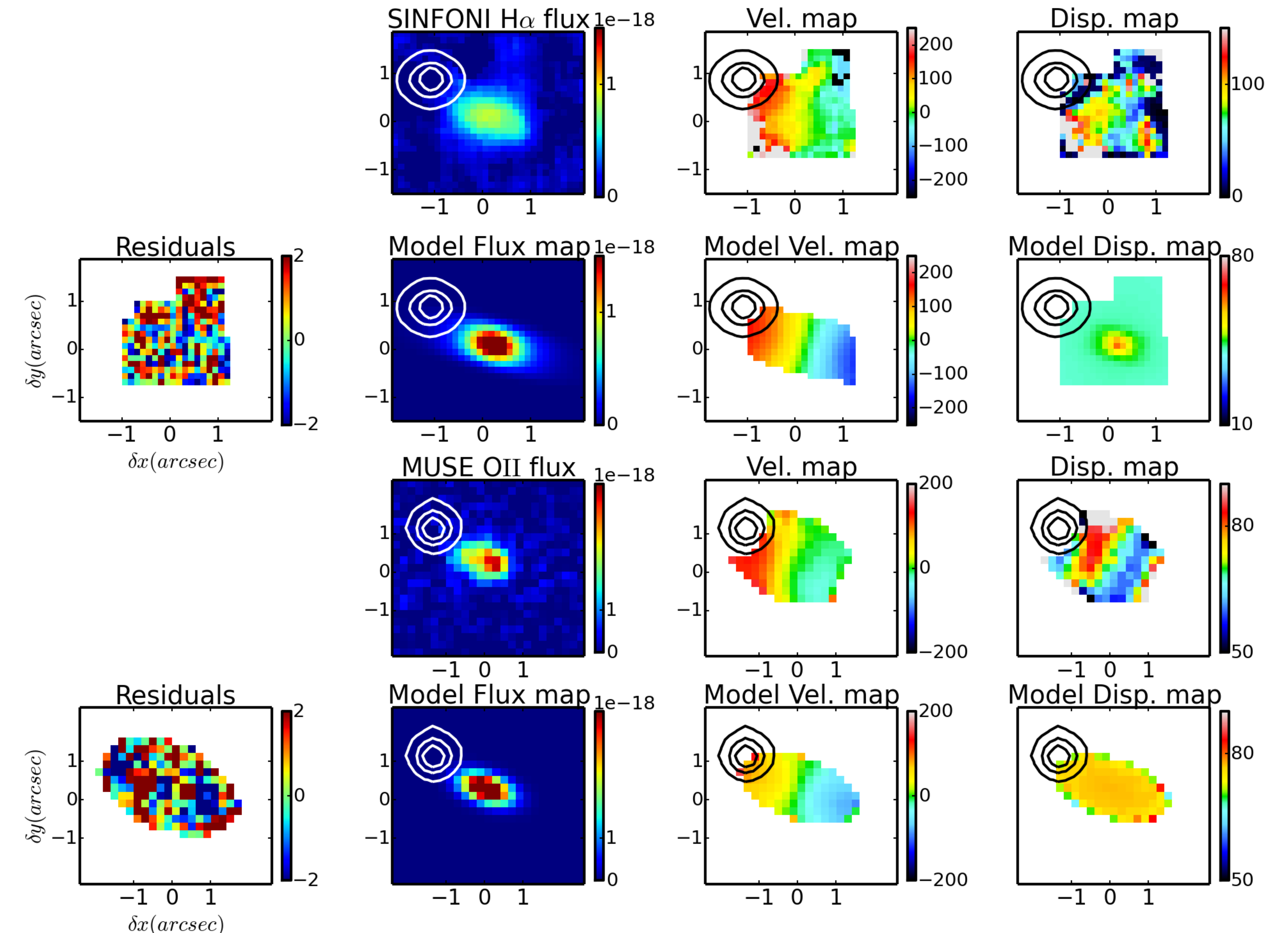}
\caption{Flux maps and kinematics from the  \Ha\ SINFONI (top rows) and the \OII\ MUSE data (bottom rows).
 The first row shows the flux (in \flux), velocity (in \kms), and dispersion (in \kms) maps from the \Ha\ SINFONI data \citep{SchroetterI_15a}.
The second row shows the residual map (in units of $\sigma$), flux, velocity and dispersion maps extracted from the 3D forward model (convolved with the PSF)
whose parameters are determined  using our \galpak\ algorithm \citep{BoucheN_15a}.
The  morpho-kinematics parameters of the model are found using the posterior distribution of the Monte Carlo Markov Chains.
The third and fourth rows show the same for the \OII\ line. 
The resulting galaxy model is determined from \Ha\ and \OII\ are similar (see Table~\ref{table:galaxy}). 
In each panel, the quasar PSF is represented by the solid contours.
}\label{fig:maps}
\end{figure*}

In Figure~\ref{fig:maps}, we compare the kinematic properties of the host derived from \Ha\ obtained with SINFONI (top two rows)
 to those derived from \OII\  obtained with MUSE (bottom two rows).
In this figure, the  flux, velocity, and dispersion maps
are determined using the 2-dimensional line-fitting algorithm LINEFIT and CAMEL for the \Ha\ and \OII\ data
that are described in \citet{CresciG_09a} and \citet{EpinatB_12a}, respectively. CAMEL allows one
to fit the two components of the \OII\ doublet simultaneously.
This figure already shows that the kinematics extracted from \Ha\ and \OII\ are consistent with each other,
and that the azimuthal angle of the quasar apparent location $\alpha$ is $\sim15^\circ$ from the galaxy P.A.


In order to derive the kinematic parameters, 
we use  the \galpak\ (v1.6.0) algorithm \citep{BoucheN_15a} to fit directly the 3D
 data  using  small subcubes around \Ha\ (SINFONI), \OII, and \Hb\ (MUSE)
 after removing the galaxy continuum emission using a linear fit.
This algorithm fits a 3D   parametric disk model   to the emission-line data cube
and returns the best-fit values for each of the parameters.
The algorithm takes  into account the PSF and the instrument line-spread function (LSF),
and thus returns the intrinsic (`deconvolved') galaxy properties, such as half-light radius ($R_{1/2}$), total flux ($f_{\rm tot}$), inclination ($i$), maximum rotation velocity ($V_{\rm max}$) and the disk velocity dispersion $(\sigma)$.  As described extensively in \citet{BoucheN_15a},
it is particularly well suited for extended objects when the size-to-seeing ratio is  $0.5$--$1.0$ or greater.

Here, since  $R_{1/2}$ is about 0\farcs5 and the seeing conditions are  $\sim$0\farcs8 for SINFONI and $\sim$0\farcs6 for MUSE,
 the size to seeing ratio  is 0.65--0.9.
Hence, because it is close to the formal margin of $\sim0.75$,
it is particularly important to compare the kinematic results of SINFONI with those obtained with MUSE at higher spatial resolution.

To model the galaxy, we  used  an exponential flux profile $I(r)$, i.e., with a S\'ersic index $n=1$, and an arctangent rotation curve $v(r)\propto \arctan(r/r_{\rm t})$
where $r_{\rm t}$ is the turnover radius. The output parameters do not change when we instead use a Gaussian flux profile.
Figure~\ref{fig:maps} shows the results of the 3D fits to the data with the \galpak\ algorithm,
where we show flux, velocity, and dispersion maps extracted from 
the modeled  data cube,  {\it convolved} with PSF and instrument resolution, for comparison purposes.
 The values for the morphological and kinematic parameters are listed in Table~\ref{table:galpak}.

From the \Ha\ SINFONI data, as described in \citet{SchroetterI_15a},
the inclination ($i$) and half-light radius $R_{1/2}$ are well constrained and found to be $i\approx60^\circ$ and $R_{1/2}\approx 4$ kpc, regardless of the choice in the S\'ersic index.

For the \OII\ MUSE data, we fit  both lines of the \OII\ doublet at once, with a unique line ratio of 0.75~\footnote{A map of the line ratios performed by the linefitting algorithm reveals that it varies slightly from 0.65 to 0.85, i.e. by no more than 15\%.}. 
We found  again the inclination ($i$) and half-light radius $R_{1/2}$ to be well constrained at $i\approx60^\circ$ and $R_{1/2}\approx 4$ kpc, regardless of the choice in the S\'ersic index $n$. In other words, the morphological parameters derived from \OII\ are in good agreement with the \Ha\ derived values.  

Regarding the kinematic parameters, we found that  the  turnover radius $r_{\rm t}$ is   degenerate with the maximum rotation velocity,  as already noted in \citet{SchroetterI_15a} from the \Ha\ data. 
Hence, we set  the  turnover radius $r_{\rm t}$  to 1.5 kpc   ($\sim$1 spaxel), to satisfy
the scaling relation between   $r_{\rm t}$ and the disk exponential $R_d$ found in local disk samples \citep{AmoriscoN_10a},
which is approximately $r_{\rm t} \simeq R_d\times0.9$. 
For fixed turnover radii $r_{\rm t}=$ 1--2 kpc, 
we found that the maximum circular velocity $V_{\rm max}$ is 100--110 \kms\ for \OII\ and around $\sim120$ \kms\ for \Ha.
The intrinsic velocity dispersion is found to be 30--40~\kms. 
The other  kinematic and morphological parameters determined from SINFONI and MUSE are listed in Table~\ref{table:galpak}.

We also applied the algorithm to the \Hb\ MUSE data (shown in Figure~\ref{fig:muse:spectra}). 
The values of the morpho-kinematic parameters derived from the \Hb\ line are somewhat different
(Table~\ref{table:galpak}) with the half-light radius $R_{1/2}$ and $V_{\rm max}$ parameters being somewhat smaller.
This is likely due to the fact that the \Hb\ line lies at 9280~\AA\ which is at the far end of the MUSE spectral range (9300~\AA) where
the throughput is much lower.  Underlying interstellar absorption in the \Hb\ profile may also play a role~\footnote{Note that \Ha\ interstellar absorption equivalent width is much smaller than that of \Hb.}.

From the galaxy size and maximum velocity, we estimate its dynamical mass within its half-light radius to be
 $M_{\rm dyn}(r<R_{1/2})\equiv R_{1/2}V_{\rm max}^2/G\approx 2\pm0.4 \times10^{10}$~\msun.
Its halo mass $M_{\rm h}$ is estimated to be $M_{\rm h}\approx 1.9\pm1.5 \times 10^{11}$ \msun\ using
\begin{equation}
M_{\rm h} \approx 2\times10^{11}\; V_{\rm max,100}^3 \;({1+z})_{1.909}^{-1.5}\;\msun
\end{equation}
where $ V_{\rm max,100}$ is the maximum rotation velocity in units of 100 \kms, and the redshift factor $1+z$
is normalized to $1.909$. The corresponding halo virial radius $R_{\rm vir}=V_{\rm max}/10/H(z)$ is 
 $R_{\rm vir}\sim 90\, V_{\rm max,110}$~kpc, assuming $V_{\rm vir}=V_{\rm max}$.

\subsection{Wind}
\label{section:wind}

In this last subsection, we return to Figure~\ref{fig:muse:spectra}, where the MUSE host galaxy  spectra showed
  self-absorbed \MgII$\lambda\lambda2796,2803$ components. 
This  \MgII$\lambda\lambda2796,2803$ component is blueshifted  
with respect to the galaxy systemic velocity, which is  most easily explained by a wind being launched from the galaxy
\citep[as in][among others]{WeinerB_09a,ErbD_12a,MartinC_12a,RubinK_14a,BordoloiR_14a,ChisholmJ_15a,HeckmanT_15a,WoodC_15a}.

Our MUSE data also show indirect indications of \MgII\ and \FeII\ emission. Indeed, the second panel of
Figure~\ref{fig:muse:spectra} shows clearly that the \MgII$\lambda2796$ optical depth is lower than the  \MgII$\lambda2803$ transition, 
a situation opposite from the expected oscillator strengths, which are 0.6 and 0.3, respectively. This is likely the result of 
\MgII\ emission infill as proposed by  \citet{ProchaskaJ_11a} and  \citet{ScarlataC_15a} and discussed in  \citet{ErbD_12a}. 
Our MUSE data also show the presence of \FeII$\lambda\lambda2587,2600$ absorption as shown in the bottom panel  in Figure~\ref{fig:muse:spectra}.
Here  the \FeII$\lambda2587$ optical depth is higher than that of the  \FeII$\lambda2600$ transition,
again opposite from the expected oscillator strengths, which are 0.07 and 0.24, respectively \citep{ErbD_12a}.
The wind scattering emission models of \citet{ProchaskaJ_11a} and \citet{ScarlataC_15a} naturally account for these two apparent anomalies,
but they   also predict prominent fluorescent emission Fe*$\lambda\lambda$2612,2626 and \MgII\ emission profiles, as seen in
 stacked spectra \citep{ErbD_12a,TangY_14a,ZhuG_15a} and in the individual cases of \citet{RubinK_11a} and \citet{MartinC_13a}.

Our MUSE data  reveal  no signs of fluorescence emission \FeII*$\lambda\lambda$2612,2626 nor \FeII*2632,
after median-smoothing or summing over large apertures.
This absence of direct fluorescent emission signatures in the presence of resonant \FeII\ and \MgII\ 
infilling can be explained  by several arguments  \citep{ProchaskaJ_11a}.
First, a large dust opacity ($\tau_{\rm dust}>10$) would suppress the emission, but this is not consistent with our
 our low reddening values ($E(B-V)<0.1$).
Second, \citet{ProchaskaJ_11a} showed that the emission is suppressed in an anisotropic wind (their Figure~9),
where the opening angle $\theta_{\rm w}$ is much smaller than 45 $^\circ$.
Lastly, the emission signal may be hidden by the galaxy continuum, and our data lack the required S/N to unveil the emission.

Another potential strong limitation of the  \citet{ProchaskaJ_11a} and \citet{ScarlataC_15a}
  models that may explain the absence of fluorescent emission originates in
the underlying assumption of a wind velocity that scales linearly with distance $v\propto r^{1}$. 
If one relaxes this assumption and
instead uses an arctangent wind velocity profile $v(r)$ motivated by the results of
\citet{MurrayN_05a,MurrayN_11a}, which 
has  a steep acceleration profile (or a large velocity gradient) inside some turnover radius,
the spatial extent of the region of resonance will be much smaller than the scale length of the wind itself.
Outside  the turnover radius, where the wind velocity is constant with radius, the
 Sobolev approximation breaks down because $\rm d v/\rm d r\approx0$, and dust re-absorption may play a much larger role in spite of low reddening values.
A full wind scattering model is beyond the scope of this paper.

We now attempt to estimate the mass flux in the wind.
In the case of a mass-conserving flow, the mass outflow rate can be estimated using this formula \citep{HeckmanT_00a,HeckmanT_15a}
\begin{eqnarray}
\dot M_{\rm in}(b) 
&\propto& \Omega_{\rm out}\,\NH\,r_{\rm out}\, V_{\rm out}\,  ,  \label{eq:wind} \\
&\approx & 0.3 \,\Omega_{\rm out, 2}\; \mu_{1.6}\;  N_{\rm H, 20.4} \; r_{\rm out, 1} \; {V_{\rm out, 80}}  \; \mpy\nn
\end{eqnarray}
where $\Omega_{\rm out}$ is the wind solid angle, $\mu$ the mean particle weight, $\NH$ the gas column density,
$r_{\rm out}$ the launch radius, and $V_{\rm out}$ the wind speed.
With blueshifted low-ionization lines in galaxy spectra, only the  wind speed is well constrained. 
As argued in \citet{BoucheN_12a} and \citet{SchroetterI_15a}, the launch radius $r_{\rm out}$ is the most uncertain ingredient of Eq.~\ref{eq:wind}.

We can estimate the wind speed $V_{\rm out}$ from the blueshifted \MgII\ absorption in the galaxy spectra.
Using a Gaussian fit to the \MgII$\lambda$2803 component (or two Gaussians to the doublet), we find a Doppler offset
of $-45\pm15$ \kms. However, the \MgII\ absorption harbors a significant component from the ISM at zero velocity. 
Hence, performing a double-Gaussian fit to the absorption representing the ISM~\footnote{Because such a double-Gaussian fit is highly degenerate,
we fix the ISM component at $v=0$ \kms, and set its width to 35 \kms, a value taken from the nebular line emissions.} and the wind components \citep[as in][]{MartinC_12a,KacprzakG_14a}
shown in Figure~\ref{fig:muse:spectra},
 we find that the wind speed (at peak optical depth) is about $V_{\rm out}\approx -80\pm15$ \kms. This is the bulk velocity where
most of the optical depth is, but the \MgII\ profile (and the \FeII profile) shows absorption up to approximately $-150$~\kms.  
For a bi-conical flow,
this wind speed represents the flow speed along any radial trajectories, but if the absorption originates from regions close to the disk
where the flow is more cylindrical,  $V_{\rm out}$ ought to be corrected for the galaxy inclination ($\cos i$) and would be -160~\kms\ since the galaxy inclination is $\sim60^{\circ}$.
 
Following \citet{LeithererC_13a},  \citet{WoodC_15a}, and  \citet{HeckmanT_15a}, one can estimate the hydrogen column density
 using the gas-to-dust ratio and the relationship between \NH\ and reddening  \citep[e.g.][]{BohlinR_78a,DiplasA_94a,MenardB_09a},
\begin{equation}
N(H)=4.9\times10^{21} \times E(B-V)\; \cmsq,
\end{equation}
which with our estimate of $E(B-V)\sim0.1$ leads to a total gas column density of $\NH\sim 5\times10^{20}$~\cmsq.
As argued in  \citet{WoodC_15a}, this value represents an upper limit on the column density, as the extinction traces the column density
to the star cluster and thus likely includes   contributions from  gas in both the disk and the wind.
Another lower limit comes from the \MgII\ rest-frame equivalent width  $W_r^{2796}\sim3.5$\AA\ (Table~\ref{table:fluxes}),
which implies a column density $>2\times10^{20}$~\cmsq\ from the \citet{MenardB_09a} column density-\MgII equivalent width correlation,
i.e.  at least about 40\%\ of the column density is in the wind.

Regarding the wind solid angle $\Omega_{\rm out}$, we use the now firmly established result that winds appear well collimated
  \citep{BordoloiR_11a,BoucheN_12a,KacprzakG_12a,MartinC_12a,RubinK_14a,BordoloiR_14a,ZhuG_14a}.
These constraints on the opening angle $\theta_{\rm out}$ show that it is on average $\approx \pm30^\circ$. 
In our galaxy, $\theta_{\rm out}$ cannot be measured directly, although
an indirect constraint comes from the emission infill discussed above, which indicates that $\theta_{\rm out}$ is much smaller than 45$^\circ$.
Overall, the total wind solid angle $\Omega_{\rm out}$ is  $\approx 2$,
for both sides of a bi-conical flow, with likely values ranging from 1.7 to 3.9 for $\theta_{\rm out}$ ranging from 30$^{\circ}$ to 45$^{\circ}$.

As mentioned,  the launch radius $r_{\rm out}$ is the most uncertain ingredient in  Equation~\ref{eq:wind} from  blueshifted absorption lines.
 Some authors take a fixed value of 5~kpc \citep[e.g.][]{RupkeD_05a,WeinerB_09a,ChisholmJ_15a}, while others assume $r_{\rm out}=2\times R_{1/2}$
 \citep[e.g.][]{HeckmanT_15a}. However, it could be 0.5 kpc or 5 kpc, as blueshifted absorption are rather unable to distinguish these possibilities.
Here we take the conservative point of view that the wind is launched not far from the disk with $r_{\rm out}$ of 1 kpc,
which is the typical thickness for high-redshift galaxies \citep[e.g.][]{ElmegreenB_06a}.

With these  assumptions, the mass outflow rate $\dot M_{\rm out}$ for the wind seen in this galaxy is about $\dot M_{\rm out}\approx 0.3$~\mpy.
A robust upper limit comes from the maximum allowed range for the column density (\NH$_{\rm max}=4.9\times10^{21}$ \cmsq),  
($\Omega_{\rm out, max}=3.9$) and $r_{\rm out, max}=5$~kpc, and  the mass outflow rate $\dot M_{\rm out}$ is at most $<6$~\mpy.
With the considerable uncertainties in the allowed values for $\theta_{\rm out}$ and $r_{\rm out}$, the mass outflow rate $\dot M_{\rm out}$ 
is most likely between 0.2 and 6 \mpy.  

\section{Properties of the Circumgalactic Medium}
\label{section:cgm}

\subsection{Line-of-sight Abundances}

\begin{table}
\caption{UVES element abundances  \label{table:abundances}}
\centering
\begin{tabular}{lccc}	
Element & Data/Method & $N_X$ & [$X/H$] \\
& & (cm$^{-2}$) & \\
\hline
\HI  & {\it HST}/COS & 20.4 $\pm$ 0.4 &   n.a. \\
\hline
\MgI & VLT/UVES &13.11$\pm$0.07 & n.a. \\
\MgII & VLT/UVES & $\geq15.65$\footnote{Limit from the \MgII\ doublet ratio $R\leq1.07$ 
following \citet{JenkinsE_96a} and \citet{WeinerB_09a} given that $W_r^{2796}=2.8~\pm0.1$~\AA\ and $W_r^{2803}=2.6\pm0.1$~\AA. } & $\leq-0.3$ $\pm$ 0.4 \\
\FeII & VLT/UVES &15.26 $\pm$ 0.03 & $-0.6$ $\pm$ 0.4 \\
\SiII &  VLT/UVES &15.57 $\pm$ 0.07 & $-0.3$ $\pm$ 0.4 \\
\ZnII & VLT/UVES &12.91 $\pm$ 0.07 & $-0.1$ $\pm$ 0.4 \\
\CrII & VLT/UVES &13.43 $\pm$ 0.04 & $-0.6$ $\pm$ 0.4 \\
\MnII & VLT/UVES &13.08 $\pm$ 0.03 & $-0.7$ $\pm$ 0.4 \\
\TiII & VLT/UVES &12.58 $\pm$ 0.05 & $-0.8$ $\pm$ 0.4 \\
\AlIII & VLT/UVES &13.53 $\pm$ 0.03 & ---\\
\hline
$\logZ$ & Jenkins09 & -0.38 $\pm$ 0.12 & ---\\
$F_\star$ & Jenkins09 & -0.35 $\pm$ 0.09 \\
$A_V$ (mag)  & Vladilo06 & 0.07 \\
$E(B-V)$ & Vladilo06 & 0.02\\
\hline
\end{tabular} 
\end{table}

\begin{figure*}
\centering
\includegraphics[width=15cm]{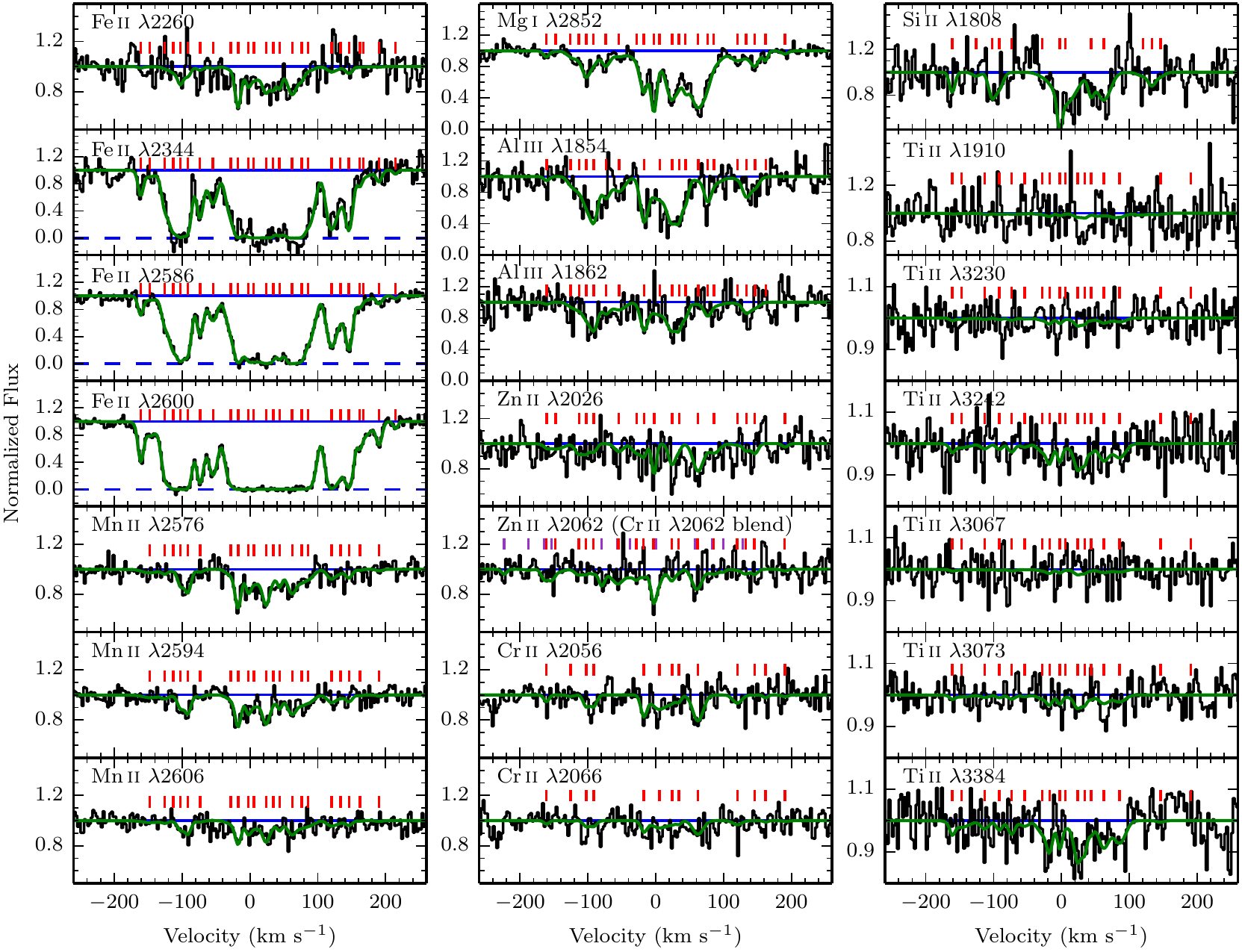}
\caption{Metal  absorption lines detected in the VLT/UVES spectrum toward the
 background quasar (solid black lines). The fit to the multicomponent
 absorption system is shown in green for each transition. Red tick marks
 indicate the position of the individual components. Note that the \ZnII$\lambda
 2062$ transition is blended with \CrII$\lambda2062$, and tick marks indicating
 components for \CrII$\lambda 2062$ are shown in orange. Zero velocity is relative
 to the galaxy systemic redshift, \redshift.
} \label{fig:uves}
\end{figure*}

We now turn to the analysis of the kinematics of the circumgalactic medium (CGM) seen in absorption against the background quasar.
From the VLT/UVES high-resolution spectra of the quasar, we constrain the abundances in several elements including \Zn, \Fe, \Si, \Cr, \Mn, and \Ti. Figure~\ref{fig:uves} shows each of these elements,
 and Table~\ref{table:abundances} summarizes our measurements. 
We used Carswell's VPFIT program  (v9.5; Carswell et al.: \url{http://www.ast.cam.ac.uk/~rfc/vpfit.html}) to perform a joint fit to all of the ions, where 
components in common  between two species have the same redshift and Doppler parameters ($b$-values),
to constrain the total column density in each element $N_X$.

\begin{figure}
\centering
\includegraphics[width=8cm]{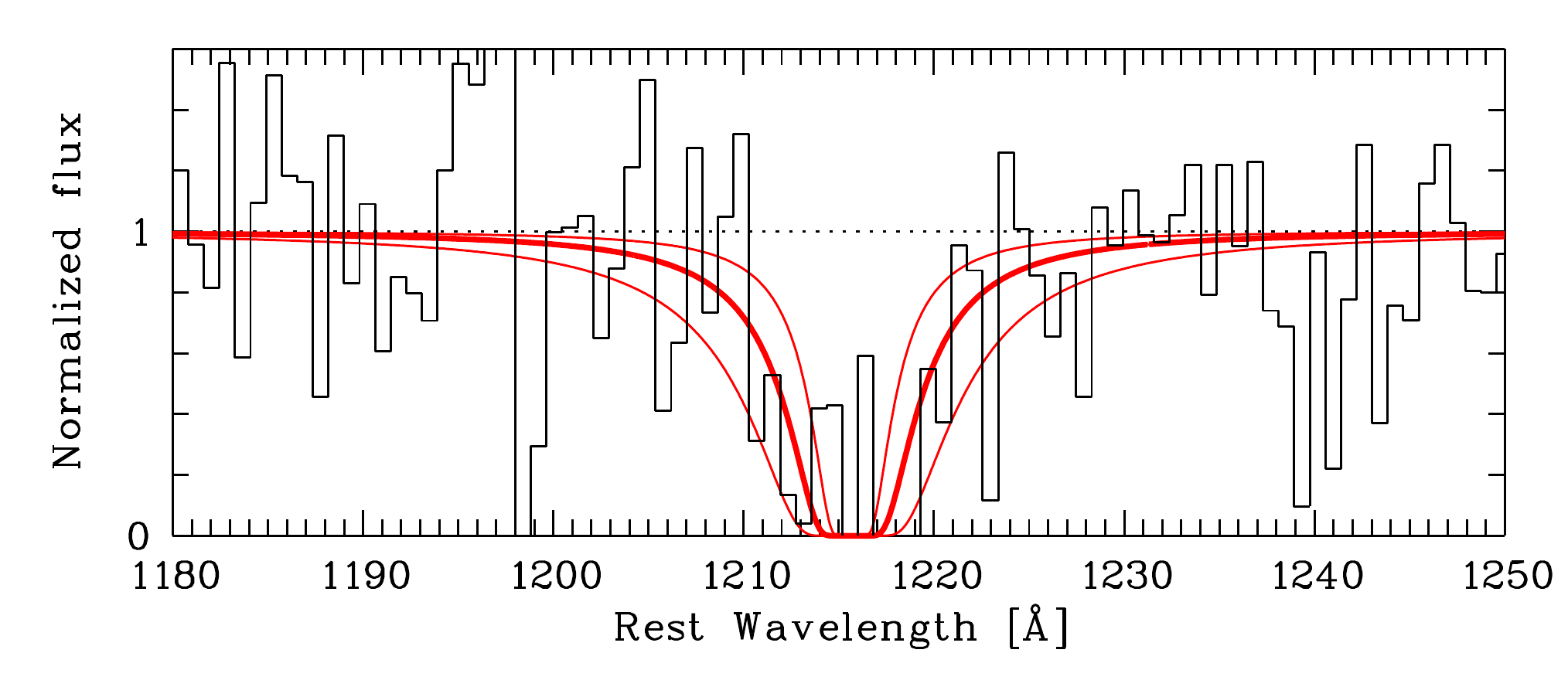}
\caption{\HI\ profile from an {\it HST}/COS G230L NUV spectrum, where the best-fit \HI\ column density is found to be $\log \NHI/\cmsq= 20.4\pm$0.4.
The shaded area represents the allowed range in \NHI. } \label{fig:cos}
\end{figure}

From our {\it HST}/COS spectra of the \Ly\ absorption, we fitted the \HI\ Voigt profile and found the column density constraint to be
$\log \NHI (\cmsq) = 20.4\pm$0.4 (Figure~\ref{fig:cos}).
Taking the \Zn\ column density and  the \HI\ column density from the COS spectra, the absorbing gas metallicity
is about $[\Zn/H]=-0.1\pm0.4$, assuming no dust depletion.
From the \Fe\ and \Zn\ column densities, we estimate the dust content in the quasar sightline to be at $A_V=0.07$
using the method proposed by \citet{VladiloG_06a}.

While \Zn\ is the least depleted element,
it may still be depleted onto dust grains. Indeed,
\citet{JenkinsE_09a} showed that in Milky Way interstellar sightlines,
 the observed ion metallicity $[X/\HH]_{\rm obs}$ of element $X$ (including \Zn)
can be described with the linear relation 
$[X/\HH]_{\rm obs}=[X/\HH]_0+A_X\;F_\star$, between   the undepleted metallicity of element $X$,  $[X/H]_0$, 
 the   propensity of that element to be depleted onto dust grains  $A_X$, and  the depletion level $F_\star$. 
\citet{JenkinsE_09a}
calibrated the propensity $A_X$s and the zero points $[X/\HH]_0$ such that the depletion level $F_\star$ {\it usually} ranges from 0 to 1  in local ISM sightlines, although some sightlines
have  negative values in regions with low gas densities $n(H)<10^{-2}$ cm$^{-3}$, as shown in their Fig. 16.
 With multiple ions of different propensity  $A_X$,  this set of linear equations (one for each element) can be solved for a unique metallicity
$Z$ and a unique $F_\star$ \citep{JenkinsE_09a}.  Furthermore, in the absence of a measurement of the $\HH$ column density, one can also
 fit simultaneously for the depletion factor $F_\star$ and for the total gas plus metal column density $\log \NHI/\cmsq + \logZ$.

\begin{figure}
\centering
\includegraphics[width=8cm]{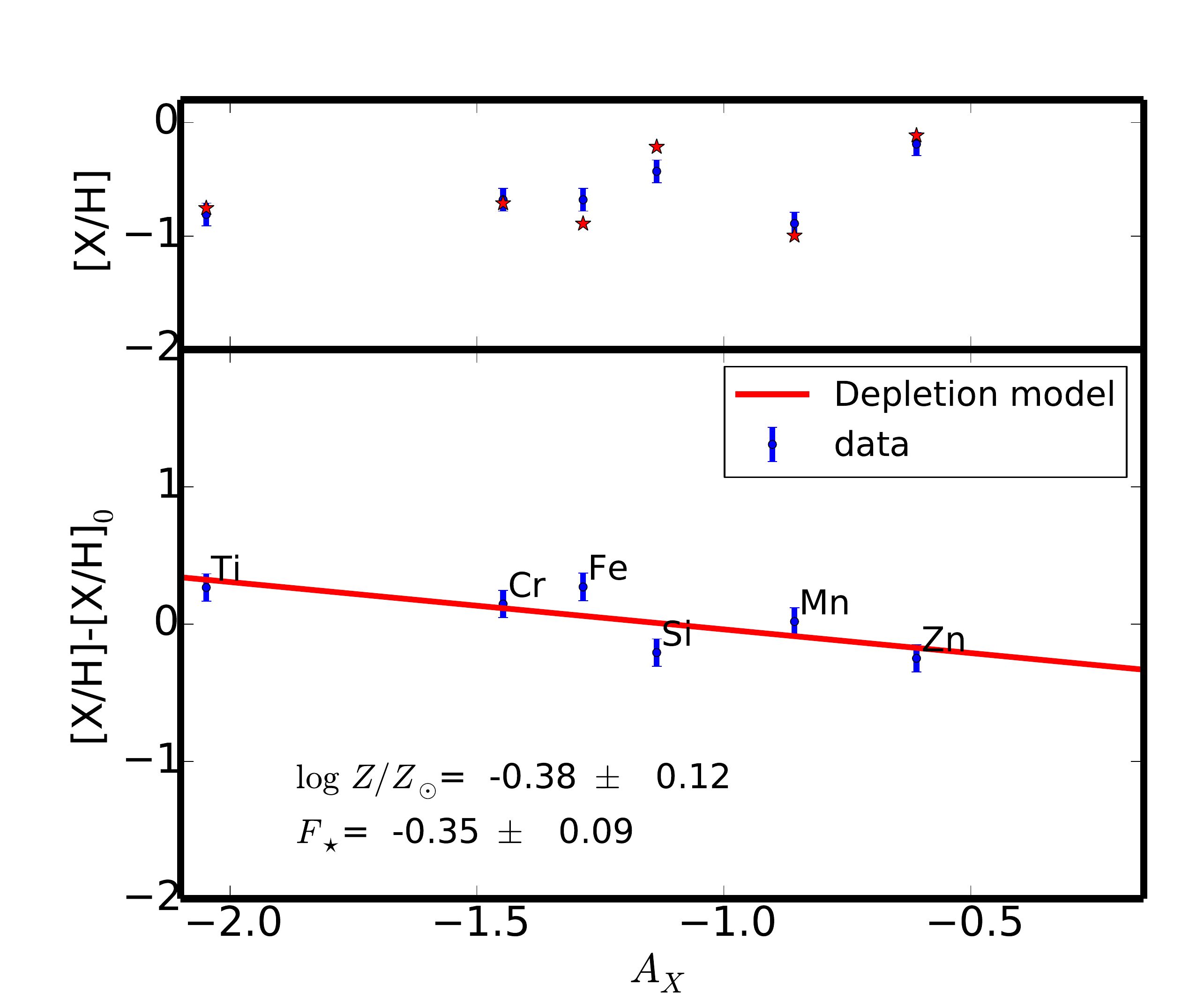}
\caption{Top:   ion abundances for the \Zn,  \Cr, \Fe, \Ti,  and \Mn\ elements present in the QSO UVES spectrum
as a function of $A_X$ (the dust propensity factor).
Bottom:   the linear fit to the global metallicity $\logZ$ where the dust depletion factor $F_\star= -0.35\pm0.1$
is given by the slope and $\logZ=-0.38\pm0.12$ is determined from the intercept using the method proposed by \citet{JenkinsE_09a}.
\label{fig:jenkins}
}
\end{figure}

Using the \citet{JenkinsE_09a} approach, Figure~\ref{fig:jenkins}
shows that a joint fit to the ion abundances (\Ti, \Cr, \Fe, \Si, \Mn, \Zn)
including the \HI\ column density of 20.4 (Table~\ref{table:abundances})
yields a metallicity of   
$\logZ = -0.38\pm0.12$ and  a low value of $F_\star =-0.35\pm0.12$, which is consistent with the low extinction value~\footnote{
As noted earlier, this negative value for the depletion level occurs in some sightlines in the Milky Way ISM in regions with gas   densities $n(H)<10^{-2}$
cm$^{-3}$ lower than the mean of the Milky-Way sample.}.
Hence, the gas metallicity probed by the background quasar at $b=12$~kpc from the galaxy is about 0.4 $Z_\odot$ ($-0.4$ dex),
albeit with large uncertainties dominated by the uncertainties on the \HI\ column density (0.4 dex).

As discussed earlier, the metallicity of the host from  the nebular lines in our MUSE and SINFONI data  is about solar ($12+\log\rm{O/H}=8.7$), implying that 
 the metallicity of the gas probed by the quasar line of sight 12 kpc away
 could be less enriched than that of the host, given the uncertainties.
More importantly, the absorbing gas appears to be highly enriched   compared to the IGM metallicity of $\logZ \simeq -2$ or less \citep[e.g.][]{SongailaA_01a,CarswellB_02a,SchayeJ_03a,AguirreA_04a,AguirreA_08a,SimcoeR_04a,PieriM_10a,PieriM_14a,ShullM_14a}, which implies
that significant mixing must have occurred with recycled gas from past outflows.

\subsection{Line-of-sight Kinematics }

We now investigate the absorption line-of-sight kinematics using the least saturated low-ionization line \MgI.
Figure~\ref{fig:mgi}(a) shows the low-ionization \MgI\ kinematic profile around the systemic velocity
for the host redshift $z_{\rm sys}=\redshift$.  The profile clearly shows a strong component  at $+$60--70~\kms, with other weaker components
at intermediate velocities, from $-50$ to $+50$ \kms. There is an additional component at $-100$~\kms.

As discussed in th next subsection,  we can gain   insights into the nature of the absorbing gas by comparing
the line-of-sight kinematics (with respect to the host galaxy kinematics) to simple models.
Such analyses are powerful, but have only been possible in very few cases,  such as in \citet{SteidelC_02a}, \citet{BoucheN_12a,BoucheN_13a}, \citet{KacprzakG_10a,KacprzakG_14a}, and \citet{SchroetterI_15a} with background quasars and in \citet{RubinK_10a} and \citet{DiamondStanicA_15a} with a bright background galaxy.
This analysis requires good constraints  on the galaxy systemic redshift and  on the galaxy's 
relative orientation with respect to the quasar sightline. Fortunately,  all of these conditions are met in this study.

\subsection{Interpretation of the Line-of-sight Kinematics }
\label{section:interpretation}

The absorption seen in the quasar line of sight shown in Figure~\ref{fig:mgi}(a) could arise in 
the following physical situations:
\begin{itemize}
\item it could be due to the low-ionization component of the outflow (scenario A);
\item it could be due to the extended parts of the ISM of the host (scenario B);
\item it could be due to infalling cold gas cooling isotropically from a hot halo (scenario C) akin to high-velocity clouds;
\item it could be due to infalling gas with significant angular momentum forming a ``cold-flow disks'' (scenario D). 
\end{itemize}
An additional scenario often invoked with quasar absorption lines is the invisible satellite possibility. This scenario can never be
ruled out in an individual quasar--galaxy pair, but from intervening \MgII\ statistics, this is the least likely possibility.
Indeed, the cross section of  satellites  is too small to account for the large $\rm d N/\rm d z$ for strong \MgII\ systems, as argued in \citet{MartinC_06a}.

Before looking at the line-of-sight kinematics, the quasar apparent location does provide tight constraints to distinguish between the
three possibilities outlined earlier.
Indeed, the quasar apparent position is located
at an azimuthal angle of only $\alpha=15^\circ$ from the galaxy major axis, i.e.
the apparent background quasar is almost perfectly aligned with the galaxy's major axis (Figures~\ref{fig:musefield:narrow} \&\ \ref{fig:vmap}).

The low azimuthal angle $\alpha$   gives a tight constraint on  the outflow scenario ``A.'' 
In order for the line of sight to intersect a bi-conical outflow,
 the outflow opening angle $\theta_{\rm out}$ ought to be much larger than $>60^{\circ}$, i.e.
be almost isotropic, given the galaxy inclination $i\sim60^\circ$. 
 This possibility is not supported by the statistical results in the literature and by our data, as discussed in Section \S~\ref{section:wind}.
Furthermore, a simple bi-conical flow model 
---which has been successful in reproducing absorption profiles in \citet{BoucheN_12a,KacprzakG_14a,SchroetterI_15a}--- 
would produce absorption at a single speed because the line-of-sight is almost entirely radial,
and thus it would  not account for 
the velocity range observed in the low-ionization profile shown in Figure~\ref{fig:mgi}(a).
Hence, we rule out the wind scenario ``A.''

The low azimuthal angle $\alpha$  also gives a tight constraint on the extended ISM scenario ``B,''
since the projection effects are minimized along the kinematic major axis. 
Figure~\ref{fig:vmap} shows that, at the  quasar location ($\Delta$R.A$=+$1\farcs38, $\Delta$decl.$=+$0\farcs84),
the projected line-of-sight velocity  is $V_z=+65$~\kms, from the modeled intrinsic (`deconvolved') velocity field
 determined by our 3D fitting algorithm \galpak.
Figure~\ref{fig:mgi}(a) shows that the maximum optical depth occurs at  $\sim60$~\kms, i.e.
is consistent with the extended parts of the ISM velocity field. However, the intermediate-velocity
components are not accounted for under this scenario.

We now investigate whether  infalling cloud scenarios, with either an isotropic or anisotropic distribution, could 
account for the intermediate-velocity components at $-50$ to $50$~\kms\ in Figure~\ref{fig:mgi}(a).
Any isotropic distribution for clouds in a galaxy halo would produce symmetric velocity distributions, which is not supported by 
our data. There are two additional arguments against this isotropic scenario. First, the \HI\ column density
 is too large ($\log N_{\HI }= {20.4}$~\cmsq) compared to the typical column density in high-velocity clouds ($\sim10^{18}$~\cmsq), as discussed in \citet{WakkerB_04a} and \citet{LehnerN_12a} and all have $\log N_{\HI}<20.2$ \citep{HerenzP_13a}.  Second, our system has
an \MgII\ rest-frame equivalent width of $W_r^{2796}\sim3.5$~\AA, whereas the typical high-velocity cloud has a 
rest-frame equivalent width of $W_r^{2796}$ of 0.3--1~\AA\ \citep{HerenzP_13a}.  
Lastly, it has been shown by several groups that the strong  \MgII\ systems with  $W_r^{2796}>0.8$~\AA\ are
not virialized in their host halo \citep[e.g.][]{BoucheN_06c,GauthierJR_09a,LundgrenB_09a}.

On the other hand, the intermediate components are qualitatively similar to the  features expected for anisotropic gas accretion
inside halos. In particular, most numerical hydro-simulations \citep{DekelA_09a,FumagalliM_11a,StewartK_11a,StewartK_11b,GoerdtT_12a,ShenS_13a}
have shown that accreting material is expected to co-rotate with the central disk in the form of a warped, extended cold gaseous 
``disk'' whose absorption kinematic signatures should follow roughly the rotation direction 
but offset  from the galaxy's systemic velocity \citep{StewartK_11b}.
Such rotating gaseous structures are found in the local universe with the large \HI\ disks 
present around diverse types of galaxies, e.g. around
the M81 massive galaxy \citep{YunM_94a},  the M33 low surface brightness disk \citep{PutmanM_09a}, and M83
\citep{HuchtmeierW_81a, BigielF_10a}, among others. Furthermore, the \HI\ column densities in the outer parts of these systems are well within the range of our observations.

In summary, the intermediate-velocity components of our absorption kinematic profile could be due to
an extended cold gaseous ``disk''  (sometimes referred to as cold-flow disk), and the largest optical depth component
at $V\sim+60$~\kms\ is likely due to the extended parts of the galaxy ISM.
We now turn towards a more detail modeling analysis of the line-of-sight kinematics.

\subsection{Line-of-sight Kinematics Model }

In order to assess whether these qualitative signatures are in agreement with the expectations 
for an  extended cold gaseous structure,
we used  a simple geometrical toy model to generate simulated absorption profiles, as in \citet{BoucheN_13a}.
In the model, we distribute ``particles''  representing gas clouds in a  co-planar structure (following the host galaxy's inclination) with
predetermined kinematics.  The model is composed of two components, one with circular orbits whose velocity is set by the galaxy rotation curve, and one with radial orbits, representing an accretion component. Because the galaxy orientation (galaxy inclination, P.A.) relative to the quasar is well determined
from the IFU data, the only free parameter is the inflow speed at the quasar impact parameter ($b=12$ kpc).

\begin{figure}
\subfigure[]{
\includegraphics[width=9cm]{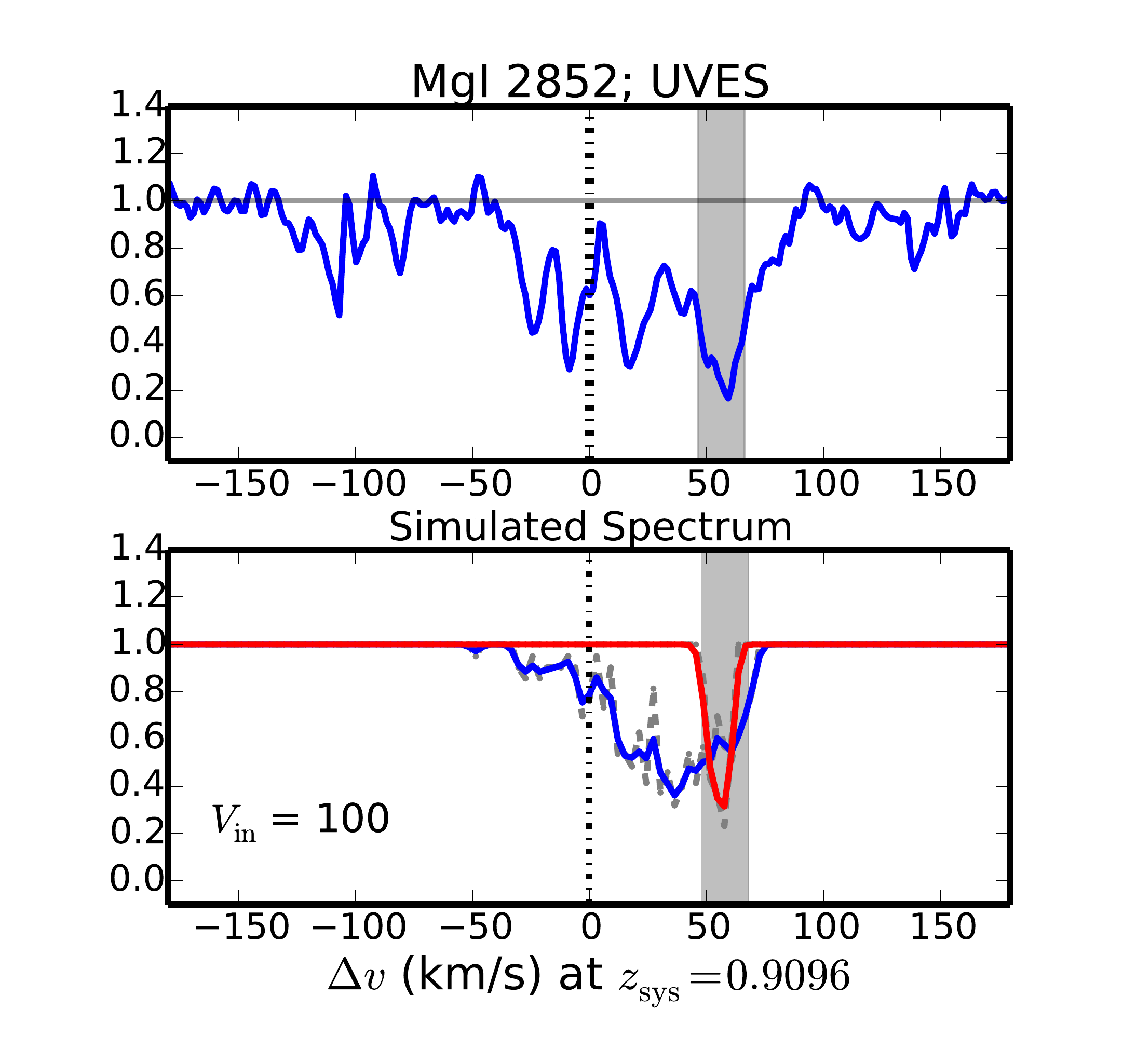}
}
\subfigure[]{
\includegraphics[width=9cm]{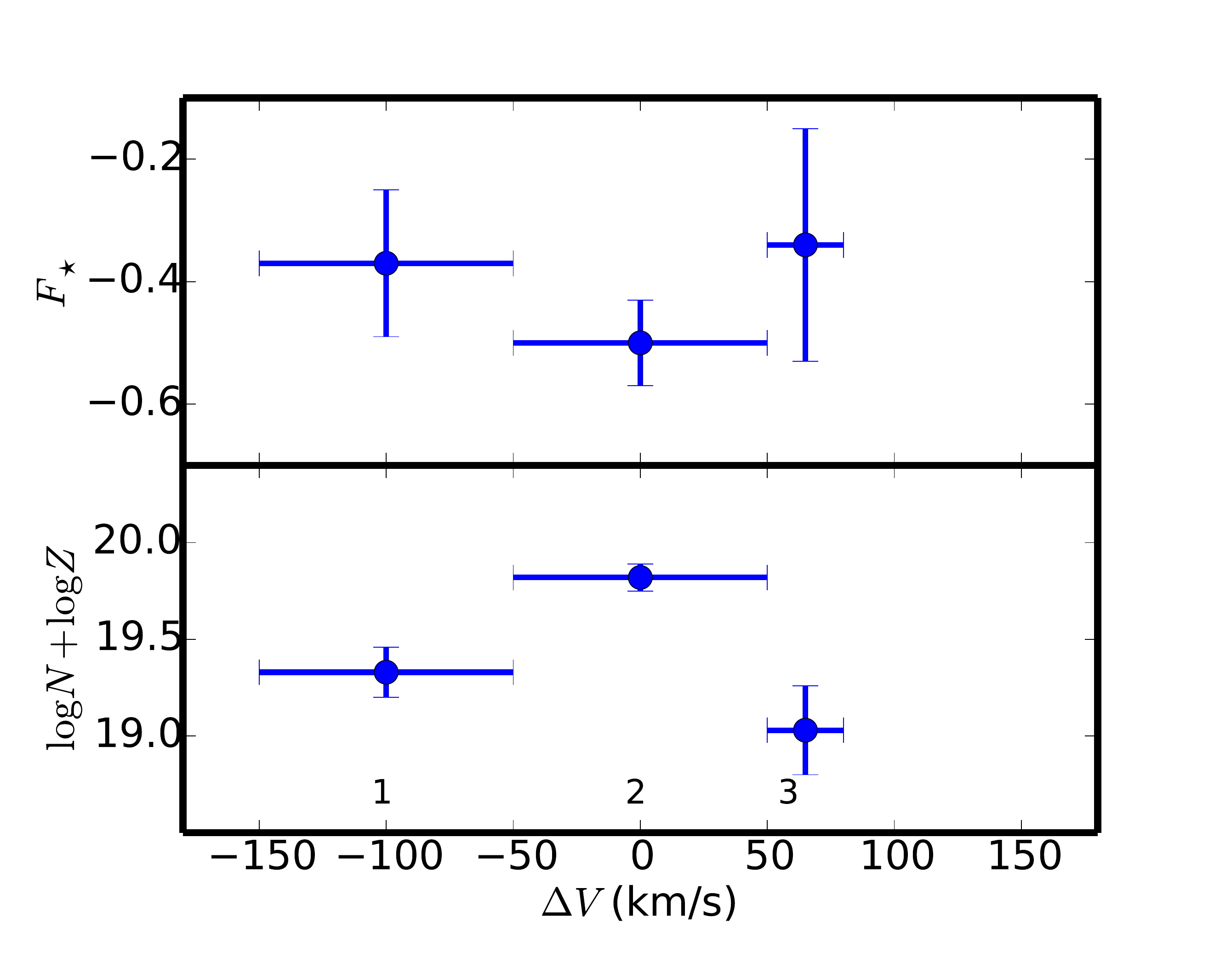}
}
\caption{(a) The top panel shows the VLT/UVES spectrum  of the \MgI\ line   as a function of the line-of-sight velocity from the systemic redshift
$z_{\rm sys}=$\redshift.
The bottom panel shows two simulated profiles (see text), one representing the absorption from a thin disk (red),
and the other representing  an inflowing model (blue) with $V_{\rm in}=100$~\kms. 
The grey band shows the expected line-of-sight velocity of $+65$ \kms\ at the quasar location determined from the velocity field shown in Figure~\ref{fig:vmap}.
There is   good qualitative agreement between the simulated spectrum and the \MgI\ profile.
(b) The depletion $F_\star$ (top) and the total column density $\log \NHI/\cmsq+\logZ$ (bottom) determined from the \ZnII,  \CrII, \FeII, \SiII\ and \MnII\ column densities in three different zones (labeled 1 to 3) showing that  region 2 has the largest column density and the lowest depletion factor $F_\star$.
 }\label{fig:mgi}
\end{figure}

\begin{figure}
\centering
\includegraphics[width=9cm]{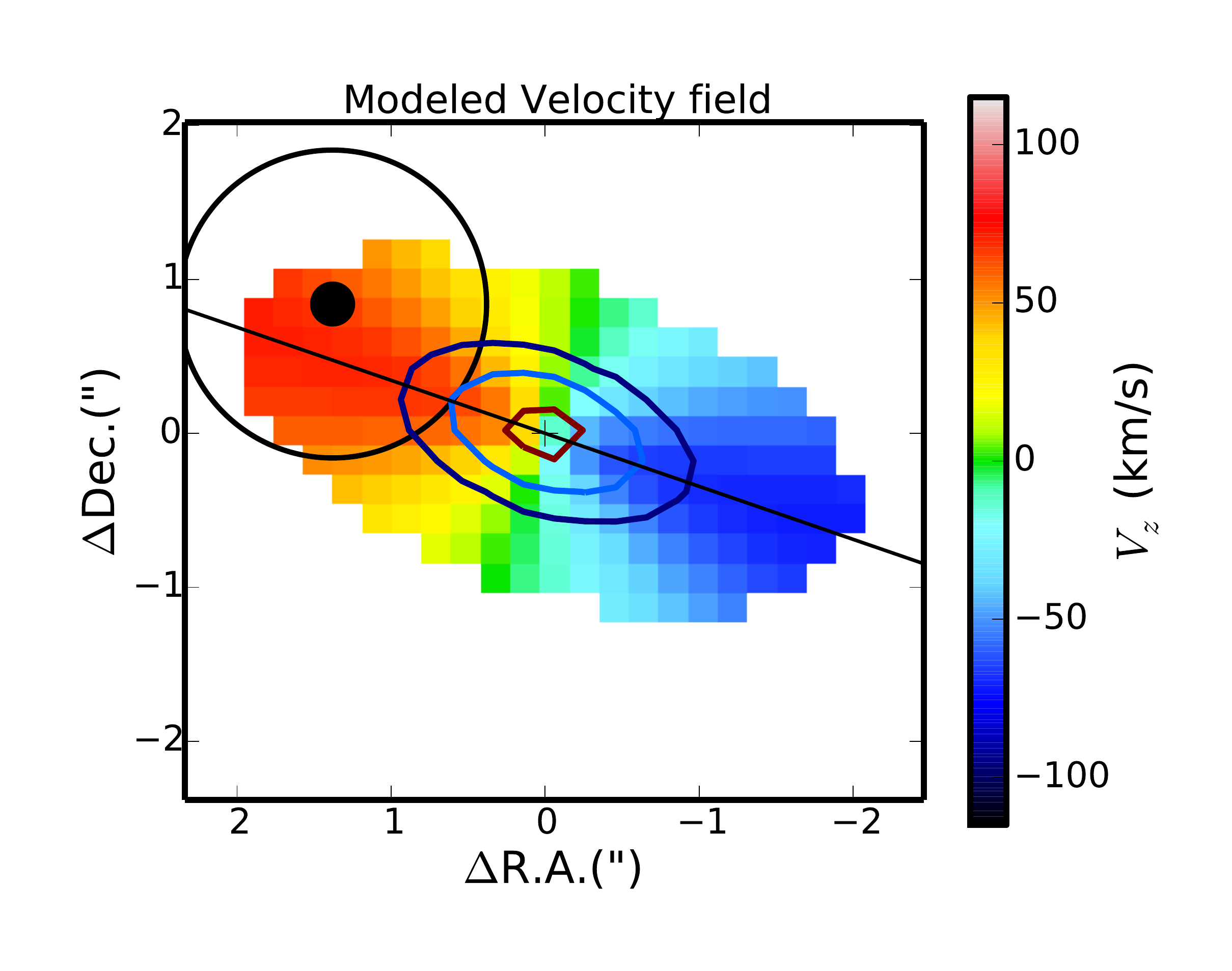}
\caption{Modeled galaxy velocity field obtained from the VLT/MUSE \OII\ data (Figure~\ref{fig:maps}). 
The contours show the intrinsic flux profile at
50, 150, and 500 $\times10^{-20}$\flux~arcsec$^{-2}$. 
The quasar location at  $\Delta\alpha=+$1\farcs38 and $\Delta\delta=+$0\farcs84 is shown by the filled black circle.
The circle of radius $r=1$" surrounding the quasar is to scale with respect to the circles shown in Figure~\ref{fig:musefield:narrow} and its insets.}
\label{fig:vmap}
\end{figure}

The resulting   absorption profile simulated at the UVES resolution is  shown 
in Figure~\ref{fig:mgi}(a) and    agrees qualitatively with the data.
The  component resulting from the galaxy's rotation is shown in red, and the component from the radial inflow is shown  
by the blue line.  
We found that an inflow speed of $\sim$ 100~\kms\ reproduces the profile shape, except for the component at $-100$~\kms,
which likely has a separate origin.
 
We can estimate the gas column density in the intermediate-velocity components using the optical depth profile of the least depleted low-ionization element \Zn, or  equivalently  using the depletion model of \citet{JenkinsE_09a} on  the \Zn,  \Cr, \Fe, \Si, and \Mn\ column densities measured in three kinematically defined sub-regions, labeled 1--3
in Figure~\ref{fig:mgi}(b). The region ``1'' is defined from the components at $\Delta V<-50$ \kms.
The region ``2'' is defined around the components at  $-50<\Delta V<50$ \kms, and region ``3'' is defined with
$50<\Delta V<80$ \kms\ corresponding to the galaxy rotation.

Figure~\ref{fig:mgi}(b) shows that the depletion factor $F_\star$ and total gas plus metal column density $\log \NHI/\cmsq + \logZ$
 are different in the three subregions.  The top panel shows that the depletion level seems the lowest in the accretion region (zone 2), consistent with this gas being the least processed.
The bottom panel shows that, provided that the metallicity does not vary across the profile significantly, about 70\%\ of the total column density
(19.90 of the total 20.05 in $\log \NHI/\cmsq + \logZ$) is carried by the middle zone ``2,'' corresponding to the accretion zone. 
With a metallicity of -0.4 dex,
the gas column density in this zone is then $\log \NHI/\cmsq \simeq20.3$. 
Note that that this 70\%\ fraction is found also using the \Zn\ column density and a uniform metallicity.

With this column density estimate of $\log \NHI/\cmsq \simeq20.2$ and  the inflow speed of $V_{\rm in}\approx$100~\kms, we can estimate the mass flux rate $M_{\rm in}(b)$ in this component from
the following arguments following \citet{BoucheN_13a}. 
For a  gaseous structure of thickness $h_z$ and mass density $\rho$ that is intercepted at the quasar impact parameter $b$,  the (radial) accretion flux $\dot M_{\rm in}$ through an area of $2\pi b\,h_z$ is  
\begin{eqnarray}
\dot M_{\rm in}(b) 
= 2\pi b\, V_{\rm in}\, \cos(i)\,m_p\mu\NH, \label{eq:accr:orig}
\end{eqnarray}\
 where $i$ is the inclination of the structure, $\NH$ is the total gas column,
$\mu$ is the mean molecular weight,
$m_p$ is the proton mass,
and we used the identity $m_p\mu\NH=\int{\rm d}z\rho(b)= \rho(b) h_z/\cos i$.  
In our case,
\begin{eqnarray}
\dot M_{\rm in}(b)  
&\geq&
8 \frac{\mu}{1.6} \;  \frac{N_{\HH}}{10^{20.3} }  \; \frac{b}{12} \;\frac {V_{\rm in}}{100}\;\frac{\cos(i)}{0.5}   \;
\mpy \label{eq:accr}
\end{eqnarray}
where $N_{\HH}$ is the gas column density (cm$^{-2}$), $b$ the quasar   impact parameter (in kpc), and ${V_{\rm in}}$ 
 the inflow velocity (in~\kms) and $i$ the galaxy inclination. 
Because  we are unable to constrain the ionization state of each of the components,  this mass flux is strictly a lower limit.


\section{Discussions}
\label{section:discussion}

\subsection{Accretion rate}

As argued in \S~\ref{section:interpretation}, the most likely physical interpretation for 
 the intermediate-velocity components in the line-of-sight absorption profile 
is that of an extended gaseous structure  co-planar with the host galaxy 
similar to several local examples, such as  M33, M81, and M83 \citep{HuchtmeierW_81a,YunM_94a,PutmanM_09a,BigielF_10a}.
In the previous section, we estimated that the minimum gas accretion rate
 given by Equation~\ref{eq:accr}   is comparable to the SFR.
In this context, it is of interest to compare this estimate to the expected theoretical gas accretion rate.
In low-mass galaxies with  halo mass below $M_{\rm h}\leq10^{12}$~\msun,
the cooling time is much smaller than the dynamical time   \citep{BirnboimY_03a,KeresD_05a,DekelA_06a},
and  the maximum theoretical gas accretion rate might be given by the growth rate of the dark matter halo $ \dot{M}_{\rm DM}$
times the baryonic fraction $f_B$ modulo an efficiency $\epsilon_{\rm in}$. 

Our galaxy toward  J1422$-$00 has a halo mass of $M_{\rm h}\approx2\times10^{11}\;V_{\rm max,110}^3$~\msun,
and using the theoretical expectation for the DM accretion rate  $\dot{M}_{\rm DM}$
from \citet{GenelS_08a} or \citet{BirnboimY_07a}, the quantity  $f_{\rm B}\times\langle \dot{M}_{\rm DM} \rangle$ is $\approx4$~\msun,
consistent with our accretion rate $\dot{M}_{\rm in}$ estimate.
Hence, our mass accretion rate estimate and the maximum theoretical gas accretion rate are comparable, which would imply that the
accretion efficiency is near 100\%, with the caveat that $\langle \dot{M}_{\rm DM} \rangle$ is a
time (or population) averaged   quantity,  which can have substantial scatter (0.3~dex) 
while  our measurement is an instantaneous quantity. 

We can look at the accretion efficiency from the point of view of the steady-state solution of the `bathtub' \citep{BoucheN_10a}
or `regulator' self-regulator model \citet{LillyS_13a} given that we have simultaneous constraints  on accretion and outflow rates. 
 As shown in \citet{BoucheN_10a} and many others since \citep{DaveR_12a,KrumholzM_12a,LillyS_13a,FeldmannR_13a,FeldmannR_15a,DekelA_14a,ForbesJ_14a,PengY_14b}, galaxies at $z>4$ can be thought of as  a simplified gas regulator where there is a balance between the SFR and the gas accretion rate. 
The  equilibrium solution  can be written as
\begin{equation}
\hbox{SFR}\approx\frac{\epsilon_{\rm in}f_{\rm B}}{1+R+\eta}\dot{M}_{\rm DM}\label{eq:epsin}
\end{equation}
where $R$ is the gas returned fraction (from massive stars) under the instantaneous recycling approximation and
$\eta$ the loading factor from stellar feedback.  Our galaxy toward  J1422$-$00  has
an SFR of \sfr\ and a mass outflow rate $M_{\rm out}\approx0.2$--6~\mpy, yielding a mass loading factor at most unity $\eta\leq1$,
and together these numbers imply that $(1+R+\eta)\times$SFR is at least (with $\eta=0$) $>6$~\mpy\ and at most (with $\eta=1$)
11~\mpy.  Together, with  the maximum theoretical accretion rate  
$f_{\rm B}\times\langle \dot{M}_{\rm DM} \rangle\approx4$~\msun,
the accretion efficiency ought to be high,  near 100\%.

 In all,  our results indicate that the accretion efficiency is high at $ \epsilon_{\rm in}\approx1.0$, regardless of the loading factor $\eta$,
in agreement with the theoretical expectation of \citet{DekelA_06a} and with the empirical measurements of \citet{BehrooziP_13a}.

\subsection{Angular Momentum under the Cold-flow Disk Interpretation}

A critical question for these  extended gaseous disks is how much angular momentum they carry as discussed in \citet{StewartK_13a} and \citet{DanovichM_15a}
compared to that of the disk. We use the common definition of the spin parameter \citep[e.g.][]{BullockJ_01a},  
\begin{equation}
\lambda\equiv \frac{j}{\sqrt{2}\,R_{\rm vir}\,V_{\rm vir}}, \label{eq:bullock}
\end{equation}
where $j$ is the specific disk angular momentum $J/M$.

As discussed in \citet{MoH_98a},   the relation between disk scale length $R_{\rm d}$ and the disk spin $\lambda$ parameter
is $R_{\rm d}\propto \lambda\, R_{\rm vir}\, H(z)^{-1}$, where the proportionality constant depends on the dark matter profile.
Using  an isothermal profile, we find $\lambda_{\rm gal}=0.047$ \cite[Eq.~12 of][]{MoH_98a}.
Using a \citet{NavarroJ_97a} (NFW) profile, we estimate $\lambda_{\rm gal}$ to be 0.04--0.05, over a wide range of the disk baryon fraction
(ranging from 0.01 to 0.1). Without direct measurement of the stellar mass,   we can use the baryonic
 Tully-Fischer relation (TFR) to estimate the baryon fraction.
Using the  baryonic TFR for intermediate redshifts galaxies \citep[][and references therein]{ContiniT_15a}, 
we find that $M_{\rm bar}\approx5\times10^{9}$~\msun, yielding a baryonic fraction of
 $M_{\rm bar}/M_{\rm h}=0.025$, i.e. two to three percent, a value consistent with the halo-abundance matching techniques \citep[e.g.][]{MosterB_10a,BehrooziP_13a}.

The angular momentum of the cold-flow disk is harder to estimate without a direct size constraint, but we can place useful limits.
Since the gas is traced 12 kpc away from the galaxy center, a distance corresponding to three times the half-light radius or $0.15\times R_{\rm vir}$,
assuming the virial relation $R_{\rm vir}\sim 90\, V_{\rm max,110}$~kpc at $z\sim1$, we find that 
the gaseous disk carries 50\%\ more angular momentum  than that of the galaxy, which has $R=4$ kpc and $V_{\rm max}=110$~\kms.
Similarly, the spin parameter of the cold-flow disk $\lambda_{\rm cfd}$ is estimated to be
  $\lambda_{\rm cfd}>0.06$ since the ratios between spin parameters and specific angular momenta are identical in a given halo (Eq.~\ref{eq:bullock}.
This limit on the cold-flow disk angular momentum is consistent with the theoretical expectation of \citet{DanovichM_15a},
where the baryons within $0.3\;R_{\rm vir}$ have 2--3 times the galaxy angular momentum.

\section{Conclusions}
\label{section:conclusions}

We presented results on a single quasar--galaxy pair toward the quasar SDSS J142253.31$-$000149 
selected from our   SIMPLE
survey  (Paper I)  which consisted of searching for the host galaxies around strong ($\EW>$2~\AA) $z\sim0.8$--1.0 \MgII\ absorbers selected from the SDSS database.
The  background quasar location is 1\farcs4 away (12 kpc) from the host and is situated about 15$^\circ$ from the galaxy's major axis.

In summary, thanks to our new  VLT/MUSE data  and our ancillary SINFONI data, 
 we found that the $z_{\rm sys}=\redshift$ host galaxy of this galaxy--quasar pair 
\begin{itemize}
\item  is isolated with no neighbors within 240 kpc down to a SFR of 0.22 \mpy\ ($5\sigma$), as shown in Figure~\ref{fig:musefield:narrow}; 
\item  has a dust-corrected SFR of \sfr, using a \citet{ChabrierG_03a} IMF, and a small amount of extinction $E(B-V)=0.1\pm0.1$ mag;
\item has  a solar metallicity ($12+\log \rm{O/H}=8.7\pm0.2$) from an analysis of the nebular emission  lines detected
 in MUSE   (\OII, \Hb) and SINFONI  (\Ha), as shown in Figure~\ref{fig:metallicity};
\item has a maximum rotation velocity of $V_{\rm max}\simeq100\pm10$~\kms, corresponding to a halo mass of $M_{\rm h} \simeq2\times10^{11}$~\msun\ or to a 0.1 $L_{\star}$ galaxy,
 and an  inclination of  about $i\approx60^\circ$ simultaneously determined through 3D modeling
(Figure~\ref{fig:maps});
\item  has a wind with an  estimated mass outflow rate of 0.5--5 \mpy\ (i.e. a loading factor $\eta\leq1$) from 
the blueshifted (by $v_{\rm out}=-80\pm15$  \kms) low-ionization absorptions (\MgII\ and \FeII) in the MUSE galaxy  spectrum (Figure~\ref{fig:muse:spectra}).
 The doublet line ratios indicate emission infilling \citep{ProchaskaJ_11a}, but the MUSE data do  not show fluorescent emission
down to $\approx5\times10^{-18}$~\flux~arcsec$^{-2}$ (3$\sigma$). 
\end{itemize}

In addition, we confirmed the SINFONI results (Paper~II; Figure~\ref{fig:vmap}) that showed that the quasar is located at an azimuthal angle of $\alpha\approx15~^\circ$ from the galaxy major axis, which makes it very well suited to investigate the presence of extended gaseous structures.
The analysis of the quasar absorption profile obtained with the  VLT/UVES spectrograph shows
\begin{itemize}
\item  distinct signatures of   co-planar gas that appears to be co-rotating with the host galaxy, but at a speed lower than the rotation velocity (Figure~\ref{fig:mgi}a)
from  the low-ionization metal absorption lines;
\item that the metallicity of the absorbing gas is estimated at about  $\logZ=-0.4~\pm~0.4$ (0.4 $Z_\odot$; Figures~\ref{fig:uves} and \ref{fig:jenkins})
 globally across the profile~\footnote{We are able to constrain the relative metallicity across the absorption profile (Figure~\ref{fig:mgi}b), but not the absolute metallicity.}, which is much larger than the IGM metallicity of $\logZ = -2$
for fresh infalls, implying a significant amount of recycling.
\end{itemize}

We discussed various interpretations of these absorption signatures (in \S~\ref{section:interpretation}) and argued that the most likely interpretation
is one that is analogous to  large \HI\ gas disks seen in the local universe, which can extend 2--3 times larger than the half-light radius.
In numerical simulations, such structures can appear at significant look-back times and are sometimes referred to as ``cold-flow disks'' \citep{StewartK_11a,StewartK_13a}.
In  this context, we estimated that the amount of infalling/accreting material is $\geq8$ \mpy\ (Eq.~\ref{eq:accr}), i.e. about two  times larger than the SFR. 
By comparing the estimate of the gas accretion rate and the expected gas inflow rate for a halo  of $2\times10^{11}$~\msun, we find that the  accretion efficiency $\epsilon_{\rm in}$ is $\approx1.0$ for wind loading factors $\eta\leq1$. 
Finally,  we find that the angular momentum of the co-planar structure
 is at least 50\%\ larger than that of the galaxy given the minimum extent of this structure.

Similar cases appeared recently in the literature. There is our work \citep{BoucheN_13a}, which showed observational signatures similar to 
ones presented here in a $z=2.3$ galaxy, and there are the recent IFU observations  of a giant \lya\ emitting filament by  \citet{MartinC_15a}
around a high-redshift quasar \citep{CantalupoS_14a} which provide possible evidence for
kinematics compatible with a larger (220 kpc in radius) gaseous rotating disk.
The similarity between the kinematics of these gaseous structures and those of some hydrodynamical simulations
\citep{StewartK_11a,ShenS_13a,DanovichM_15a} and the evidences provided in the current study suggest  that these structures 
  may not be uncommon in the high-redshift Universe.


\acknowledgments
We thank the anonymous referee for their comments that led to an improved manuscript.
We thank E. Emsellem for his insights regarding some of the figures.
We thank A. Dekel for his comments on an early version of the draft.
 This work is based on observations taken at ESO/VLT in Paranal, and
we would like to thank the ESO staff for their assistance and support during the MUSE commissioning campaigns.
N.B. acknowledges support from a Career Integration Grant (CIG) (PCIG11-GA-2012-321702) within the 7th European Community Framework Program.
M.T.M. thanks the Australian Research Council for \textsl{Discovery Project} grant DP130100568, which supported this work.
J.S. acknowledges from the European Research Council (ERC) under
the European Union's Seventh Framework Program (FP7/2007-2013) / ERC Grant agreement 278594-GasAroundGalaxies.
JR acknowledges support from the ERC starting grant CALENDS.
B.E. acknowledges financial support from ``Programme National de Cosmologie 
and Galaxies'' (PNCG) of CNRS/INSU, France.
Support for program 12522 was provided by NASA through a grant from the Space Telescope Science Institute, which is operated by the Association of Universities for Research in Astronomy, Inc., under NASA contract NAS 5-26555.
This work has been carried out thanks to the support of the ANR FOGHAR (ANR-13-BS05-0010-02), the OCEVU Labex (ANR-11-LABX-0060), and the A*MIDEX project (ANR-11-IDEX-0001-02) funded by the ``Investissements d'Avenir" French government program managed by the ANR.
This research made use of Astropy, a community-developed core 
PYTHON package for astronomy (Astropy Collaboration et al. 2013), NumPy and SciPy (Oliphant 2007), Matplotlib (Hunter 2007), IPython (Perez \&\ Granger 2007), and of  NASA’s Astrophysics Data System Bibliographic Services.



\end{document}